\documentclass[twocolumn]{aa}
\usepackage{graphicx}
\usepackage{txfonts}
\usepackage{aalongtable}
\usepackage{natbib}
\begin{document}

   \title{L-band (3.5 $\mu$m) IR-excess in massive star formation}

   \subtitle{II. RCW 57/NGC 3576\thanks{Table~\ref{photres} is only available
in electronic form at the CDS via anonymous ftp to cdsarc.u-strasbg.fr (130.79.128.5)
or via http://cdsweb.u-strasbg.fr/cgi-bin/qcat?J/A+A/}}

   \author{M. Maercker,
          \inst{1,2}
          M. G. Burton\inst{2}
	  \and
	  C. M. Wright\inst{3}}

   \institute{Stockholm Observatory, AlbaNova University Center,\\
	     106 91 Stockholm, Sweden\\
             \email{maercker@astro.su.se}
         \and
             School of Physics, University of New South Wales,\\
              Sydney, NSW 2052, Australia\\
              \email{mgb@phys.unsw.edu.au}
	 \and
   School of Physical, Environmental and Mathematical Sciences, University of New South Wales@ADFA\\
   Canberra, ACT 2600, Australia\\
\email{c.wright@adfa.edu.au}}

   \date{Received October 18, 2005; accepted December 7, 2005}

   \abstract{We present a JHK$_s$L survey of the massive star forming region RCW 57 (NGC 3576) 
based on L-band data at 3.5 $\mu$m taken with SPIREX (South Pole Infrared Explorer), and 
2MASS JHK$_s$ data at 1.25-2.2 $\mu$m. This is the second of two papers, the first one 
concerning a similar JHK$_s$L survey of 30 Doradus. Colour-colour and colour-magnitude diagrams are used to detect sources with infrared excess. This excess emission is interpreted as coming from circumstellar disks, and hence gives the
cluster disk fraction (CDF). Based on the CDF and the age of RCW 57, it is possible to draw
conclusions on the formation and early evolution of massive stars. The infrared excess is detected by comparing the locations of sources in JHK$_s$L colour-colour and L vs. (K$_s$-L) colour-magnitude diagrams to the reddening band due to interstellar extinction. A total of 251 sources were detected. More than 50\% of the 209 sources included in the diagrams have an infrared excess.
Comparison with other JHK$_s$L surveys, including the results on 30 Doradus from the first paper, support a very 
high initial disk fraction ($>$80\%) even for massive stars, although there is an indication
of a possible faster evolution of circumstellar disks around high mass stars. 33 sources 
only found in the L-band indicate the presence of heavily embedded, massive Class I protostars.
We also report the detection of diffuse PAHs emission throughout the RCW 57 region. 
   
   }

   \maketitle

\section{Introduction}
\label{intro}
\subsection{IR-excess as a measure of circumstellar disks}
\label{intro1}
This paper is the second of two papers using IR-excess in the JHK$_s$L plane (1.2 - 3.5 $\mu$m) to measure 
the fraction of sources with circumstellar disks in high mass star forming regions. The first paper 
concerned 30 Doradus in the LMC (Maercker \& Burton \cite{maerckerburton}, from now referred to as 
Paper I). IR-excess can be detected using near infrared colour-colour diagrams by comparing the position 
of sources relative to the reddening vectors due to interstellar extinction. The excess radiation above 
that of a blackbody can be explained by models of circumstellar disks around a young stellar object
 (eg. Lada \& Adams \cite{ladaadams}). Although this excess radiation can be detected using JHK$_s$ data (1.2 - 2.2 
$\mu$m) alone, the nature of IR-excess is not always clear. JHK$_s$L observations give a larger separation to the IR-excess sources in colour-colour diagrams, whereas JHK$_s$ observations tend to underestimate the fraction of stars with IR-excess. On account of the difficulties of ground based observations at longer wavelengths, the L-band (3.5 $\mu$m) proves to be the best wavelength for detecting circumstellar disks, although longer wavelengths are preferable, if space observations are available. Kenyon \& Hartmann (\cite{kenyonhartmann}) show the advantage of (K$_s$-L) as a measure of IR-excess by comparing
the frequency distributions in young stellar clusters for (H-K$_s$) and (K$_s$-L). Whereas the (H-K$_s$) distribution has
one clear peak at (H-K$_s$)$\sim$0.2-0.4 and a long tail, the (K$_s$-L) distribution has a clear second peak at (K$_s$-L)$\sim$0.8-1.0 made up mostly of class II sources with optically thick, circumstellar disks. 

\subsection{RCW 57}
\label{rcw57}
This young massive star forming region, also known as NGC 3576, is one of  the brightest HII regions in 
the infrared in our Galaxy. The kinematic distance is 3.0 $\pm$ 0.3 kpc, adopted from De Pree et al. 
(\cite{depreeetal}). An asymmetrical structure in the region can be seen in the 21 cm map by Retallack 
\& Goss (\cite{retallackgoss}), which extends to the northeast but has a sharp cut off
in the southwest. The spectral energy distributions of five objects detected using
a 10 $\mu$m map (Frogel \& Persson~\cite{frogelpersson}) suggest that these are protostellar objects 
with silicate absorption features, therefore indicating Class I objects (Persi et al.~\cite{persietal}).
Near infrared photometry by Persi et al., together with an 8 - 13 $\mu$m CVF spectrum of IRS 1 (IRAS 
11097-6102 in the IRAS Point Source Catalogue), show that the majority of stars ($>$ 70\%) have an 
infrared excess in the JHK$_s$ plane. 19 of these sources could be matched with the present data, 15 of which 
we show can also be classified as having an infrared excess in the JHK$_s$L plane. The 
sources discussed by Persi et al. are confined to the central region, confirming the youth of the 
cluster. Radio recombination lines were detected by McGee \& Gardner (\cite{mcgeegardner}), Wilson et al. 
(\cite{wilsonetal}) and De Pree et al. (\cite{depreeetal}). The detection of maser sources in CH$_3$OH and 
H$_2$O (Caswell et al.~\cite{caswelletal95} and~\cite{caswelletal89} respectively) are indications of early 
stages of star formation in a dense circumstellar environment. Thorough investigations of the central 
region have been undertaken by Figuer\^edo et al. (\cite{figueredoetal}) in the near-infrared (NIR), and 
Barbosa et al. (\cite{barbosaetal}) in the mid-infrared (MIR). Nine of the MIR sources match sources seen
in the L-band image presented here. In the NIR, JHK$_s$ colour-colour and 
colour-magnitude diagrams show sources affected by excess emission, indicating the presence of circumstellar 
disks around the less massive members of the cluster (Figuer\^edo et al.~\cite{figueredoetal}). Eight spectra 
of the brightest sources show rising continua towards the IR. Three of these have a clear infrared excess. 
The detection of CO bandheads (2.2935$\mu$m) in emission and absorption indicates the presence of several sources still heavily embedded in their stellar birthclouds (Figuer\^edo et al.~\cite{figueredoetal}, Barbosa 
et al. \cite{barbosaetal}). Based on the radio data (Goss \& Shaver~\cite{gossshaver}), RCW 57 can be classified as a Giant HII (GHII) region, with ${1.6}\times{10}^{50}$ photons $s^{-1}$ in the UV (defining sources brighter than $10^{50}$ Lyman continuum photons per second as GHII regions (Figuer\^edo et al.~\cite{figueredoetal})). A possible ionizing source has been found at the peak emission of the 3.4 cm map (DePree et al.~\cite{depreeetal}), the source being a cluster of stars that have broken out of their natal cocoons but remain hidden behind dark clouds along the line of sight (Barbosa et al.~\cite{barbosaetal}). The radio peak emission (at $\sim$ RA 11h11m51s, Dec -61\degr18\arcmin45\arcsec (J2000)) is also hidden behind clouds in the SPIREX image. This is
further confirmed by Walsh et al (\cite{walshetal}), who find the peak of the 8.64 GHz continuum emission to lie
at approximately the same position, behind clouds in the N-band image (their Fig. 2). In their follow up survey, Barbosa et al. for the first time resolved IRS 1 into four sources in the 10 $\mu$m band, 
approximately 1.5\arcsec apart from each other (Nos. 48, 50, 60 and 60b in their paper. Source numbers labelled
in Table~\ref{photres} with  `No' and a number are from Barbosa et al.). One of these shows evidence for a UC HII region and they conclude that the sources in the central region of RCW 57 are in the UC HII region phase. The position of IRS 1 coincides with the brightest L-band source in our study (m$_L$=4.1, \#88, Table~\ref{photres}) and is also found in the L-band by Moneti (\cite{moneti}) with magnitude m$_L$=4.05. Barbosa et al. also found a new MIR source, without a counterpart in the NIR, which is possibly a hot core. In the central region a strong CO $J=2-1$ line at 230 GHz was 
observed by White and Phillips (\cite{whitephillips}). Shock-excited H$_2$ line emission may also indicate 
the presence of gas outflows (Figuer\^edo et al.~\cite{figueredoetal}).

\section{Observations}
\label{observations}
\subsection{L-band data from SPIREX}
The data at 3.5 $\mu$m was taken with the 60 cm South Pole InfraRed Explorer (SPIREX) (Hereld~\cite{hereld};
Burton et al.~\cite{burtonetal}) at the Amundsen-Scott South-Pole Station using the Abu camera equipped 
with a 1024x1024 InSb array (Fowler et al.~\cite{fowleretal}) in July and August 1999. 
Observations were carried out using the L-band filter (${\lambda}_{central}$=3.514 ~$\mu$m, 
$\Delta\lambda$=0.618~$\mu$m) and had a field of view of 10\arcmin and pixel scale 0.6\arcsec. The image is 
a mosaic of a series of 3 minute frames, each shifted by 15\arcsec~from each other, interleaved with
seperate sky observations, and comprises a total of 60 minutes of on-source integration. Reductions were 
done automatically using the SPIREX/Abu pipeline\footnote{http://pipe.cis.rit.edu}. Unfortunately the data 
was not flux calibrated and seperate observations were necessary to determine the calibration, as 
discussed in Paper I (see \S~\ref{calcasp}).

\subsection{Narrow Band Filters from SPIREX}

In addition to the L-band data, images were obtained through three
narrow band filters, centred on the wavelengths of the H$_2$ v=1-0
Q-branch lines, a PAHs emission feature, and the hydrogen Br $\alpha$
line. The line centres of the filters were 2.42, 3.30 and 4.05 $\mu$m,
respectively, and their widths 0.034, 0.074 and 0.054 $\mu$m.  For the H$_2$
and PAHs filters the data were self-calibrated using interpolated
values for magnitudes of the sources IRS1, Persi 43 and Persi 106
(stars 88, 74 and 124, respectively in Table~\ref{photres}) from those we determined in this
Table for the K$_s$ and L bands.  For the Br $\alpha$ filter, we
self-calibrated based on a flux for IRS1 of 16 Jy at 4$\mu$m, as measured
by McGregor et al. (\cite{mcgregoretal2}).

\subsection{JHK$_s$-band data from 2MASS}
The L-band observations were complemented with JHK$_s$ data from the 2MASS point source catalogue (PSC)
(Cutri et al.~\cite{cutrietal}) and atlas images\footnote{Available 
at http://www.ipac.caltech.edu/applications/2MASS/IM/}. The 2MASS telescopes 
(Kleinmann et al.~\cite{kleinmannetal}) scanned the sky in both hemispheres in three near infrared filters 
(J, H and K$_s$; 1.25, 1.65 and 2.2 $\mu$m respectively) with limiting magnitudes of $m_J$=15.8, $m_H$=15.1 
and $m_K$=14.3. The L-band data was first matched with sources in the PSC. Photometry was also performed on 
the K$_s$-band atlas images to derive the magnitudes for any sources that could not be matched with those in the 
PSC.

\subsection{Calibration data from CASPIR}
\label{calcasp}
Since the SPIREX L-band image was not flux calibrated, additional observations were carried out using the 
Australian National University (ANU) 2.3 m telescope at Siding Spring Observatory, equipped with CASPIR
(Cryogenic Array Spectrometer/Imager)(McGregor~\cite{mcgregor}). The observations were carried out in
early April 2004 together with the calibration observations for 30 Doradus (Paper I). A narrow band 
filter (${\lambda}_{central}=3.592 ~\mu$m, $\Delta\lambda=0.078~\mu$m) was used to avoid saturation due to 
sky brightness, in contrast to the broader band it was possible to use with SPIREX because of the lower sky background at the South Pole. To see whether this introduced an error, the calibration stars were examined for colour variations by comparing the L-band to PAH flux ratios. These were found to be constant within the measurement uncertainties, making it possible to use the narrow band calibration data. The standard star used to calibrate the images is listed in Table~\ref{standards} and the stars in the SPIREX image used to determine the zero-point correction are listed in Table~\ref{cali}. The individual error for the calibrated stars is $\sim$0.05 mags which leads to a weighted error in the 
zero-point correction of $\sim$0.03 mags and is included in all subsequent error calculations.

\begin{table}
\caption{Standard star used to calibrate the CASPIR images (McGregor~\cite{mcgregor}).}
\label{standards}
\centering
\begin{tabular}{c c c c}
\hline\hline
name & RA (J2000) & DEC(J2000) & $m_L$\\
 & (h m s) & (d m s) & Mag\\
\hline
BS4638 & 12 11 39.1 & -52 22 06 & 4.501\\
\hline
\end{tabular}
\end{table}

\begin{table}
\caption{Stars in RCW 57 used for calibration. Bright, isolated stars were 
chosen from the SPIREX image and used to calibrate the remaining stars in the 
image. Their L-band magnitudes were determined from the standard 
star, and have an average error of $\pm 0.05$ mags.}
\label{cali}
\centering
\begin{tabular}{c c c c}
\hline\hline
id & RA (J2000) & DEC(J2000) & $m_{L}$\\
 & (h m s) & (d m s) & Mag\\
\hline
8   & 11 11 10.6 & -61 17 45.2 & 7.4 \\
19  & 11 11 19.7 & -61 15 26.0 & 7.2 \\
173 & 11 12 25.4 & -61 15 11.6 & 7.6 \\
\hline
\end{tabular}
\end{table}

\section{Results}
\label{results}

   \begin{figure}
   \centering
   \includegraphics[width=8.8cm]{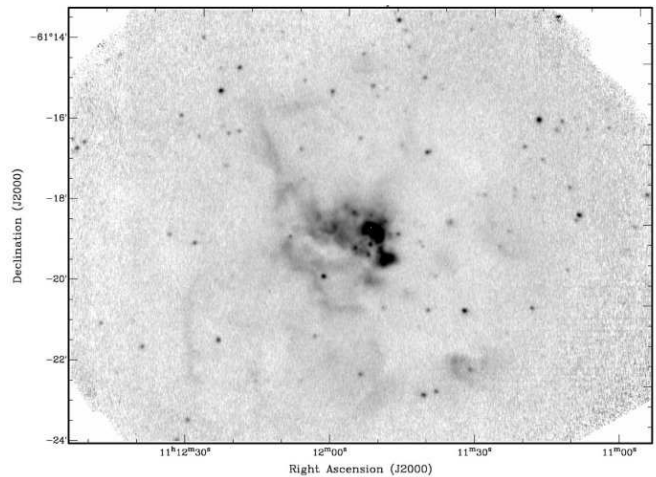}
   \caption{SPIREX L-band (3.5 $\mu$m) image of RCW 57. Total on-source integration of 60 minutes;
effective resolution 2.6$\arcsec$; pixel scale 0.6$\arcsec$; 90\% completeness limit at 11.2 mag;
faintest star detected 13.5 mag.}
              \label{rcw57im}
    \end{figure}

\begin{figure}
\centering
\includegraphics[width=8.8cm]{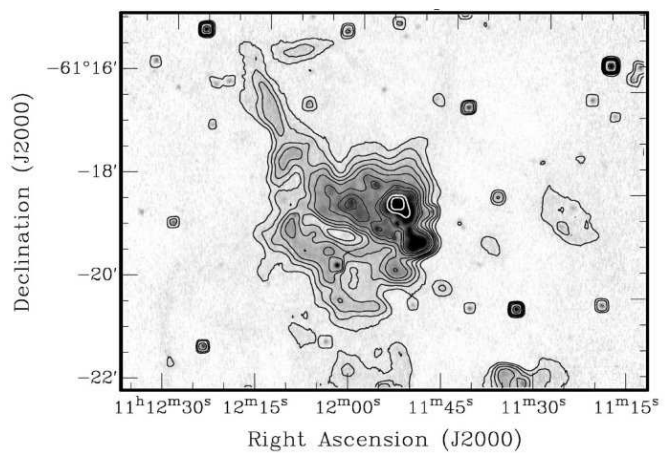}
\caption{Enlargement of the central region of RCW 57 at L-band (3.5 $\mu$m) with contours showing the
nebulosity. Contour levels are 0.03, 0.2, 0.4, 0.7, 1.0, 1.7, 2.3, 3.0, 8.3 and 20 mJy/arcsec$^2$.}
\label{rcw57cont}
\end{figure}

\begin{figure}
\centering
\includegraphics[width=8.8cm]{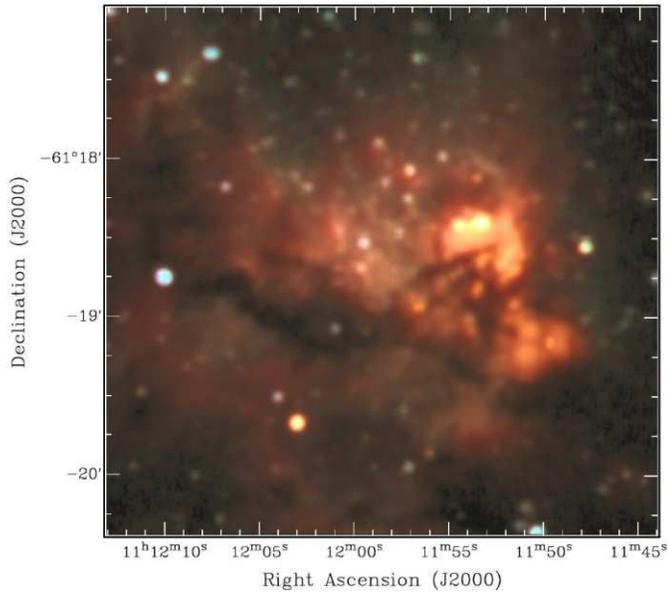}
\caption{HK$_s$L (Blue=H, Green=K$_s$, Red=L) composite colour image of RCW 57 created using 2MASS and SPIREX
images. Regions bright in the L-band (3.5$\mu$m) can be seen in red or orange, indicating the presence of young stellar objects. Prominent dust lanes are also apparent.}
\label{colour}
\end{figure}

   \begin{figure}
   \centering
   \includegraphics[width=8.8cm]{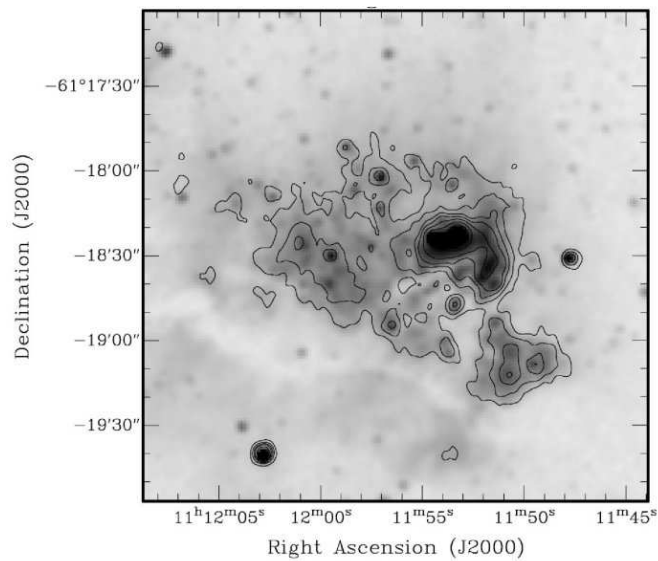}
   \caption{Contour map of the emission measured through the H$_2$ Q-branch 2.42$\mu$m
narrow band filter, overlaid on the 2MASS image of the K$_s$ band emission
from the central of RCW 57.  Contours levels are at 0.8, 1.5, 2.2,
2.9, 4.3, 7.2 and 13 $\times$ 10$^{-4}$ Jy~arcsec$^{-2}$.}
              \label{kh2}
    \end{figure}

   \begin{figure}
   \centering
   \includegraphics[width=8.8cm]{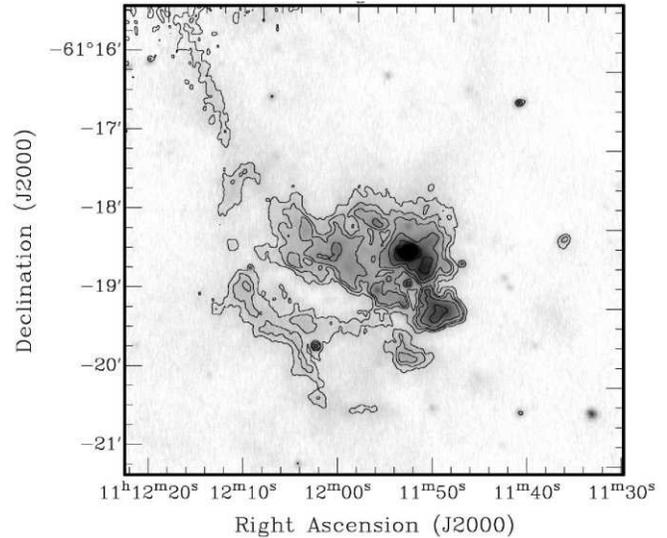}
   \caption{Contour map of the emission measured through the PAHs 3.3$\mu$m narrow
band filter, overlaid on the L-band image of RCW 57.  Contours levels
are at 2.2, 3.9, 5.7, 10 and 19 $\times$ 10$^{-3}$ Jy~arcsec$^{-2}$ for continuum, or
equivalently 4.4, 8.0, 12, 20 and 38 $\times$ 10$^{-17}$ W m$^{-2}$ arcsec$^{-2}$ if it is line
emission.}
              \label{lpah}
    \end{figure}

   \begin{figure}
   \centering
   \includegraphics[width=8.8cm]{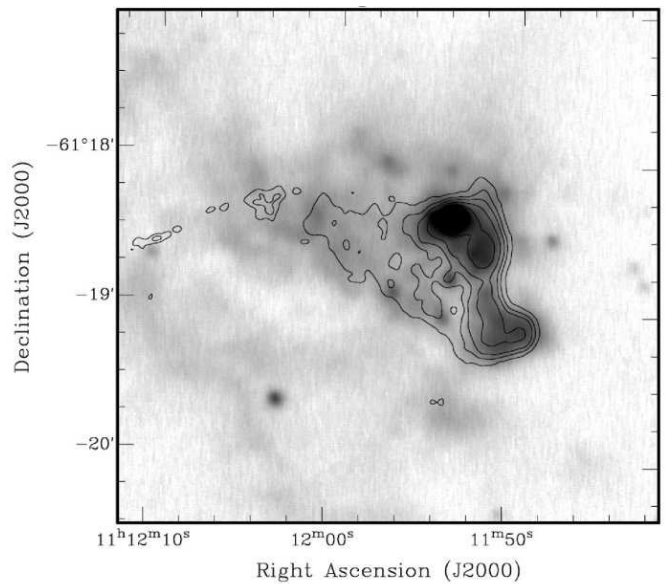}
   \caption{Contour maps of the emission measured through the Br $\alpha$ 4.05$\mu$m
narrow band filter, overlaid on the L band image of the central region
of RCW 57.  Contour levels are at 3.2, 4.9, 6.5, 9.8, 16, 23 and 61 $\times$
10$^{-2}$ Jy~arcsec$^{-2}$.}
              \label{lbra}
    \end{figure}

\subsection{Photometry}
The L-band image of RCW 57 from SPIREX is shown in Figure~\ref{rcw57im} and an enlargement of the central
region overlaid with contour lines showing the nebulosity is shown in Figure~\ref{rcw57cont}. The same 
steps to obtaining the photometry were undertaken as described for 30 Doradus in Paper I, including fitting
coordinates using the \emph{Karma} package\footnote{http://www.atnf.csiro.au/karma/}, running
\emph{IRAF/daophot} to get the photometry and adding 400 artificial stars using \emph{addstar} to estimate
the errors. This last step resulted in a 90\% completeness limit of 11.2 mag in the L-band. The individual errors for the L-band magnitudes were given by \emph{IRAF/daophot}. Matching sources
not in the 2MASS  PSC catalogue with photometry performed on the 2MASS K$_s$-band atlas images resulted in an
additional 17 matches.

The mosaiced SPIREX image suffered from irregular distortions (in the upper most left part of the image a recognizable pattern of 3 stars was 8\arcsec off). These were however small over the largest part of the image, and the L-band data could carefully be matched with the PSC and K$_s$-band images by marking the sources in the images and then selecting matches `by hand', taking local distortions into account. Table~\ref{phot} lists the statistics for detections in the various bands including the statistics when aplying the 90\% completeness 
limit.

\subsection{Sensitivity}
\label{sensitivity}
The detection threshold was taken to be three times the standard deviation $\sigma_{sky}$ of a typical 
region of sky near each source. This resulted in a limit of $\sim$12 mag at L-band, corresponding to the 84\%
completeness limit. The faintest source detected has a magnitude of $m_L$=13.5 (69\% completeness
limit). However, only sources above the 90\% completeness limit at $m_L$=11.2 mag are included in the 
determination of the IR-excess fraction (\S~4.2). 

\subsection{Foreground contamination}
\label{foreground}
Unfortunately off-source comparison images were not available for the SPIREX data. We therefore estimated
the contamination by foreground stars using the (J-K$_s$) colours of the stars, to determine a limiting (J-K$_s$) 
colour due to interstellar reddening. Assuming that sources that are part of RCW 57 are additionally 
reddened, excluding all sources bluer than the (J-K$_s$) limit gives a first order estimation of the number of
foreground stars. Figuer\^edo et al. (\cite{figueredoetal}) use the star HD 97499 to determine the 
interstellar component of the reddening to RCW 57. This results in an extinction parameter
of $A_K$=0.43 which in turn gives a limiting (J-K$_s$) colour of 0.7. This limit corresponds also to a gap
in the J vs (J-K$_s$) diagram for RCW 57, indicating main sequence stars between the source and Earth on
the blue side of the gap and the stars belonging to RCW 57 or further away on the right side of the 0.7
(J-K$_s$) colour limit. Applying this limit to the sources detected in RCW57 results in 17 potential foreground 
stars. Of these, 3 have moderate IR-excess ((K$_s$-L)$<$2) and 7 have larger excess ((K$_s$-L)$\sim$3-5). The extreme red colours in potential foreground stars is suprising. However, a closer inspection of the K$_s$- and L-band images shows that the 2MASS sources possibly are foreground objects close to the position of embedded stars (on the sky), and therefore result in large (K$_s$-L) colours. In two cases (\# 33 and \#43), the J-band magnitude is only the 95\% upper confidence limit (the uncertainty in the magnitude is given as `null' in the PSC, indicating that the source either was not detected in J-band or is inconsistently deblended), resulting in an uncertain (J-K$_s$) colour. Excluding these potential foreground stars decreases the disk fraction from 55\% to 54\% (see \S~\ref{diskevo}). Likely foreground stars are indicated in Table~\ref{photres} and are marked with boxes in Figs.~\ref{jhkl} to~\ref{cmd}.

\subsection{Narrow Band Filters}

The three narrow band images are shown in Figs.~\ref{kh2},~\ref{lpah} and~\ref{lbra}. 
In all three images the contours showing the emission through the line
filters are similar to the nearby continuum images they are overlaid
on.  Indeed, there is no clear evidence for line emission in either
the H$_2$ or Br $\alpha$ images, although this does not preclude their
presence.  

Oliva \& Moorwood (\cite{olivamoorwood}) report a detection of 2.6 $\times$
10$^{-13}$ erg cm$^{-2}$ s$^{-1}$
through a 30\arcsec~ beam centred on RCW 57 for the H$_2$ Q-branch emission,
with the flux falling to approximately one tenth that value a beam
away.  Their measurements were through a CVF of similar spectral
resolution to the narrow band filter we employed, with the line being
approximately 10\% of the continuum.  This would not be apparent in our image
without careful continuum subtraction from the narrow band image.
While H$_2$ emission is indeed likely present in RCW 57, all we can
conclude is that it is not distinguishable from the continuum in a
narrow band filter image.  There are no extended regions of bright
pure H$_2$ line emission evident.

Similarly, we cannot distinguish any Br $\alpha$ emission from the
continuum at 4$\mu$m in a narrow band filter, although again this line is
most likely present.  Moorwood \& Salinari (\cite{moorwoodsalinari}) report a 
detection of Br $\alpha$ 10\arcsec S of IRS1, but at a level that is about 10\% of the
continuum level we measure.  They also report a detection of Br $\alpha$
on IRS1 itself, though at a level inconsistent with an upper limit of
McGregor et al. (\cite{mcgregoretal2}).  In either case, the level would not be
distinguishable from the strong continuum from IRS1 in a narrow band
filter.

We can, however, report a clear detection of PAHs emission from RCW
57.  While the diffuse emission in the PAHs and L band images are
similar in morphology, with extended emission around IRS1 and a
filament extending from it to the NE, the contrast of the diffuse
emission to the stellar sources is clearer through the PAHs filter
than through the L band filter.  A quantitative comparison suggests
that typically about one third of the diffuse flux measured in
the L band filter must come from the PAHs feature at 3.3$\mu$m.  It is
likely that scattered continuum dominates the remaining diffuse
emission at L band, although we cannot demonstrate that from the data
here.  The PAHs emission is presumably fluorescently excited by the
far-UV photons also generated by the ionizing source(s) which excites
the HII region. As can be seen in our Figure~\ref{lpah}, typical fluxes for the
3.3$\mu$m PAHs emission are around 10$^{-16}$ W m$^{-2}$ arcsec$^{-2}$ in the nebulosity.

\section{Analysis}
\label{analysis}
\begin{table}
\caption{Number of detections in the different bands. The first column gives the total
number of detections in the SPIREX image. The second column gives the number of stars
that could be matched with the 2MASS PSC. Column three lists the number of stars 
additionally matched by comparison of the K$_s$- and L-band images. The last column lists the 
number of stars only found in the SPIREX L-band image. The second row lists the respective numbers
for stars brighter than the 90\% completeness limit. Using the (J-K$_s$) colour limit determined
in \S~\ref{foreground} suggests 17 of the stars detected at JHK$_s$L are likely foreground stars. }
\label{phot}
\centering
\begin{tabular}{c c c c c}
\hline\hline
& Total & JHK$_s$L & K$_s$L & L\\
\hline
all stars & 251 & 201 & 17 & 33\\
$m_L<$11.2 (90\% limit) & 209 & 168 & 8 & 33\\
\hline
\end{tabular}
\end{table}

\subsection{Colour-colour and colour-magnitude diagrams}
Figs.~\ref{jhkl} to~\ref{cmd} show the colour-colour and colour-magnitude diagrams. Fig.~\ref{jhkl}
shows  the (J-H) colour vs. (K$_s$-L) colour, Fig.~\ref{jhk} the (J-H) vs. (H-K$_s$) colours and Fig.~\ref{cmd}
the L-band magnitude vs. (K$_s$-L) colour. The diagrams were created using the JHK$_s$ data from 2MASS and the 
L-band data from the SPIREX image. Only sources above the 90\% completeness limit were included. The main 
sequence (thick solid line) for spectral types O6-8 to M5 and the giant branch (thin solid line) for 
spectral types K0 to M5 are plotted in each diagram. The sequences were plotted using their intrinsic 
colours (Koorneef~\cite{koorneef}) and their absolute visual magnitudes (Allen~\cite{allen}). Reddening 
vectors up to an extinction of $A_V$=30 mags are plotted as dashed lines, assuming an extinction law 
$\propto {\lambda}^{-1.7}$. The distance modulus used was 12.4 magnitudes, corresponding to the distance 
of 3 kpc to RCW 57. The crosses in the lower right corners indicate the mean errors for the stars in each figure.

\subsection{Fraction of reddened sources}
\label{irfrac}
The fraction of reddened sources was determined taking the individual errors for each star into account. 
Stars which lie at least 1$\sigma$ error to the right and below the reddening band are counted as having
an IR-excess and are marked with the star symbol. IR-excess stars in the JHK$_s$L plane are marked with the same
symbol in Figs.~\ref{jhk} and~\ref{cmd} as well. In Figure~\ref{cmd}, stars that are additionally only 
detected in K$_s$ and L (diamond shaped symbols) are included. Stars only detected in the L-band can be included 
by providing a lower limit for the (K$_s$-L) colour using the 2MASS sensitivity limit at K$_s$ (14.3 mag). These are 
indicated as circles and lie to the right of their positions in the colour-magnitude diagram. By comparing
the location of the stars in Fig.~\ref{cmd} to the location of already identified IR-excess sources (star 
shaped symbols), it is possible to also estimate which of the stars seen only at K$_s$ and L also have an 
IR-excess. Of the 8 stars that lie above the 90\% completeness limit, 6 occupy this region and so are 
counted as having an IR-excess. The 33 stars only detected in the L-band also lie in the region occupied by 
the IR-excess stars and therefore all of these are counted as having an IR-excess. The statistics are listed 
in Table~\ref{IRexcess}. The uncertainty in the number of stars with IR-excess is determined by 
counting the number of IR-excess sources when assuming a 2$\sigma$ distance to the reddening band. The
variation in this number gives an estimate of the uncertainty in IR-excess sources. This procedure excludes
four sources from the JHK$_s$L data set. The sources only detected at K$_s$ and L, and only at L, are not affected by 
this consideration. Thus, the determined fraction of IR-excess sources is 55$\pm$2\%. The JHK$_s$ data alone 
(Fig.~\ref{jhk}), would yield only 25 IR-excess sources, compared to 75 excess sources in the JHK$_s$L diagram. 
This would lead to a considerable underestimation of the CDF. Our JHK$_s$ diagram looks similar to the one 
presented by Figuer\^edo et al. (\cite{figueredoetal}) (their Fig. 4). The majority of sources lie along the
reddening vector, with several sources displaying clear IR-excess at (H-K$_s$)$>$1.5. Previous determinations
of the fraction of sources with IR-excess in JHK$_s$ colour-colour diagrams (Persi et al.~\cite{persietal}), 
show that $>$70\% of the sources have significant IR-excess. This is much higher than the fraction of excess
sources determined here. However, the region covered in Persi et al. is only 340$\times$340 square arc seconds,
compared to a radius of $\sim$440 arc seconds for the region covered in this paper. If an area equal to the 
region in Persi et al. ($\sim$3$\arcmin$ in radius) is considered, the fraction of IR-excess sources 
increases to 79$\pm$2\% (Table~\ref{region} and \S~\ref{diskevo}). The determination of sources not part of RCW 57 is only a first order estimate (\S~\ref{foreground}). In particular, the estimate is a lower limit on the CDF since it may not exclude all potential background stars. Contamination from background stars is likely to be more significant in the outer regions of the image where these are not hidden behind the molecular cloud, so resulting in a lower CDF.

\subsection{Luminosity Function}
\label{mass}
Figs.~\ref{luma} and~\ref{lumir} show the luminosity function for all sources in RCW 57, and IR-excess sources, 
respectively. Vertical lines indicate the 90\% completeness limit. Both diagrams 
cover the range from $m_L$=7.5-13.5 and peak at $m_L\sim$10. The distribution is somewhat higher and 
narrower when only taking IR-excess sources into account (Fig.~\ref{lumir}). Comparing the L-band luminosity
function with the L-band magnitudes of main sequence stars, the diagrams cover the range from spectral types
A3 to O5 and peak at early type B stars. Using the spectral types, it is possible to make crude estimates of 
the stellar masses. For early type B stars the masses lie at approximately 7-17 M$_{\sun}$ 
(Allen~\cite{allen}). Both diagrams span a mass range of $\sim$2-60 M$_{\sun}$, confirming that RCW 57 is a high mass star forming region. However, the IR-excess might severely bias the determination of
spectral types towards earlier type stars and higher masses, and so the results here can only be taken as indicative. There is a possible turnover at the 90\% confidence limit in the L-band luminosity function 
(Fig.~\ref{luma}). A turnover is however not seen in the K$_s$-band luminosity function (KLF) 
(Figuer\^edo et al.~\cite{figueredoetal}) and an IMF derived from the KLF gives a slope of $\Gamma=-1.62$, 
which is consistent with the Salpeter value (Salpeter~\cite{salpeter}). The KLF derived in Figuer\^edo et al.
included correction for non-cluster members, interstellar reddening, excess emission and photometric
completeness. The resulting cluster mass integrated from their derived IMF, is 
M$_{cluster}$ = 5.4 $\times$ 10$^3$M$_{\sun}$ (Figuer\^edo et al.~\cite{figueredoetal}). This is an upper 
limit, as their IMF is likely to be overestimated due to excess emission. The IMF derived from the L-band
luminosity function in Fig.~\ref{luma} has a slope of $\Gamma=-1.42$. The integrated cluster mass for
stars above the 90\% completeness limit is $\sim$10$^4$M$_{\sun}$ (using M$_{lower}$=5.8 M$_{\sun}$ and 
M$_{upper}$=100 M$_{\sun}$). However, the cluster mass is
dominated by contribution from stars below the completeness limit, and so is highly sensitive to the derived
value for the slope of the IMF.

\begin{table}
\caption{Number of stars found with IR-excess and the reddening fraction. Numbers in
the JHK$_s$L, K$_s$L and L columns are the number of stars with IR-excess found in the
respective bands. These are listed  assuming 1$\sigma$ and 2$\sigma$ distances from the 
reddening band. Column 5 gives the total number of sources with IR-excess at 
1$\sigma$ and 2$\sigma$. Column 6 gives the cluster disk fraction. Excluding possible
foreground stars decreases the CDF to $\sim54\%$. Only sources that are brighter than the
90\% completeness limit (11.2 mag in L-band) are including in calculating the fraction of
IR-excess sources.}
\label{IRexcess}
\centering
\begin{tabular}{c c c c c c}
\hline\hline
distance & JHK$_s$L & K$_s$L & L & total & frac\\
\hline
1$\sigma$ & 75 & 6 & 33 & 114 & $55\pm0.5$\\
2$\sigma$ & 71 & 6 & 33 & 110 & $53\pm0.5$\\
\hline
\end{tabular}
\end{table}

   \begin{figure}
   \centering
   \includegraphics[width=8.8cm]{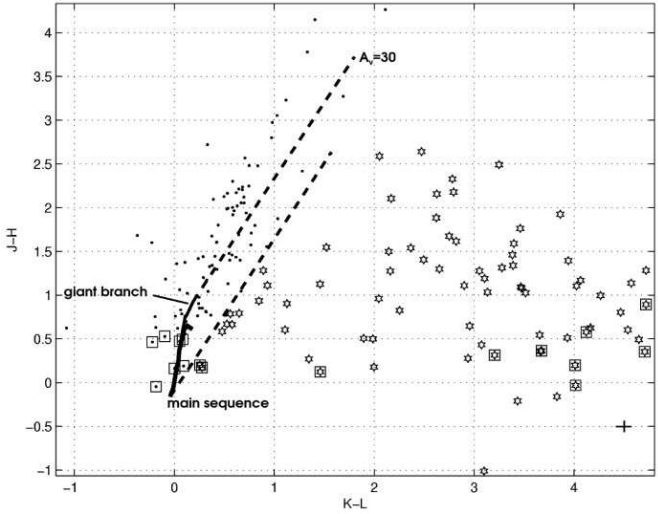}
   \caption{JHK$_s$L colour-colour diagram for RCW 57. The thick solid line shows the position of the unreddened
main sequence for spectral types O6-8 to M5. The thin solid line above shows the giant branch for spectral
types K0 to M5 and extends from (J-H)=0.5, (K$_s$-L)=0.07 to (J-H)=0.9 to (K$_s$-L)=0.19. The dashed lines show the
reddening vector up to $A_V=30$ assuming an extinction law $\propto \lambda^{-1.7}$. Star shaped symbols are
stars identified as having an IR-excess (\S~\ref{irfrac}). Sources in squares are have (J-K$_s$)$<$0.7 (see 
\S~\ref{foreground}). The cross in the lower right of the diagram indicates the mean errors for all 
stars. 209 stars lie above the 90\% completeness limit and 75 of these lie outside the reddening band and 
are therefore considered to have IR-excess (Table~\ref{phot} and~\ref{IRexcess}).}
              \label{jhkl}
    \end{figure}

   \begin{figure}
   \centering
   \includegraphics[width=8.8cm]{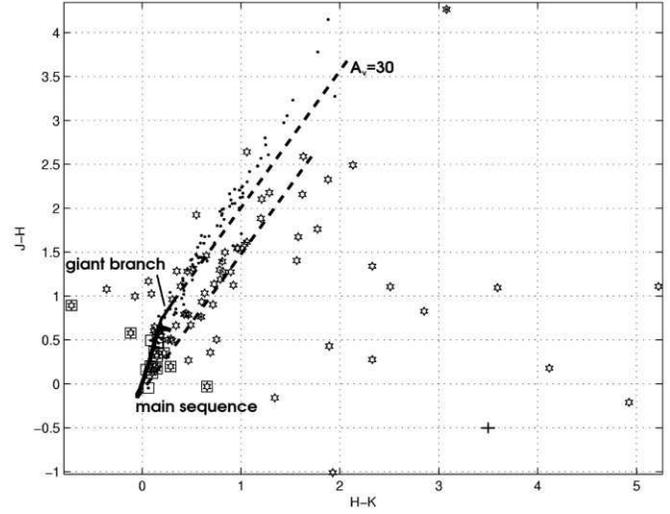}
   \caption{JHK$_s$ colour-colour diagram for RCW 57. The same symbols as in Figure~\ref{jhkl} are used. The
giant branch extends from (H-K$_s$)=0.13 to 0.31 with the same values for (J-H) as in the previous figure.
The cross in the lower right corner indicates the mean error for all stars. The
diagram shows the same sources as in the JHK$_s$L diagram using only the JHK$_s$-band data. Here the stars are 
clearly less separated than sources with IR-excess compared to using the L-band data. In this diagram only 25 sources would be classified as having IR-excess, comapred to 75 in Figure~\ref{jhkl}, leading to an underestimate of the cluster disk fraction.}
              \label{jhk}
    \end{figure}

   \begin{figure}
   \centering
   \includegraphics[width=8.8cm]{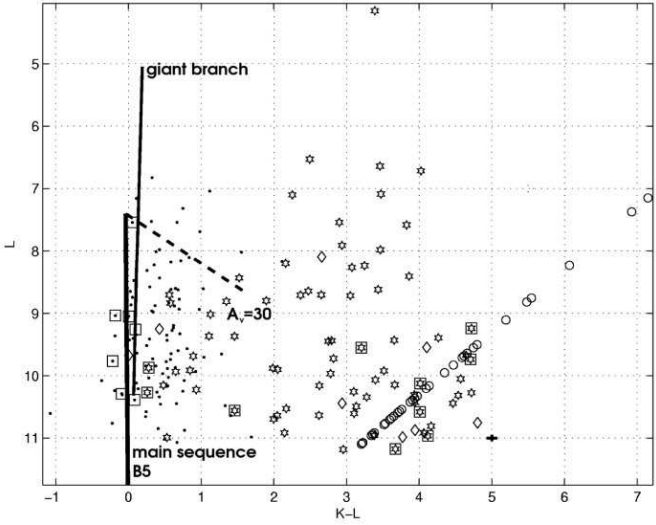}
   \caption{Infrared L vs (K$_s$-L) colour-magnitude diagram for RCW 57. The thick solid line shows the main
sequence for spectral types B5 and earlier, the thin solid line the giant branch for the same spectral 
types as in the other diagrams. The dashed line shows the reddening vector up to $A_V$=30. Star shaped 
symbols are the same IR-excess sources as in Figure~\ref{jhkl}. Diamond shaped symbols are stars only 
detected in K$_s$- and L-band and circles show the lower (K$_s$-L) limit for stars only detected in L-band
(\S~\ref{irfrac}). Squares indicate possible foreground stars with (J-K$_s$)$<$0.7; \S~~\ref{foreground}.
The cross in the lower right corner indicates the mean error for all stars. 
6 of the 8 stars detected in K$_s$ and L occupy the same region as the stars with IR-excess from Figure~\ref{jhkl}.
All 33 stars detected only in L-band are also in the region. These stars are also classified as IR-excess
sources and are counted towards the total disk fraction (Table~\ref{IRexcess}).}
              \label{cmd}
    \end{figure}

   \begin{figure}
   \centering
   \includegraphics[width=8.8cm]{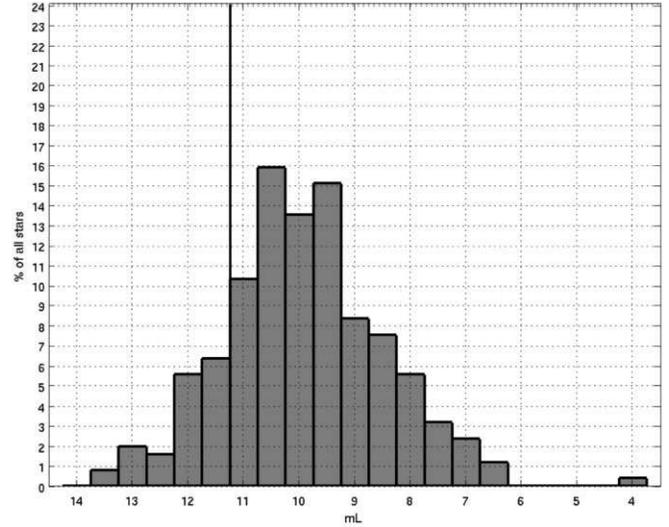}
   \caption{L band luminosity function of all stars detected at L-band in the SPIREX image of 
RCW 57. Stars are detected down to m$_L\sim$13.5, corresponding to unreddened A5 stars at the distance of
RCW 57. The upper limit is m$_L\sim$4. O5 type and earlier stars correspond to an L-band magnitude 
m$_L\sim$7.5.  The distribution peaks at m$_L\sim$10 indicating early type B
stars. Based on the spectral types the mass range is from 2-60 M$_{\sun}$ and peaks at 7-17 M$_{\sun}$. The 
vertical line  shows the 90\% completeness limit.}
              \label{luma}
    \end{figure}

   \begin{figure}
   \centering
   \includegraphics[width=8.8cm]{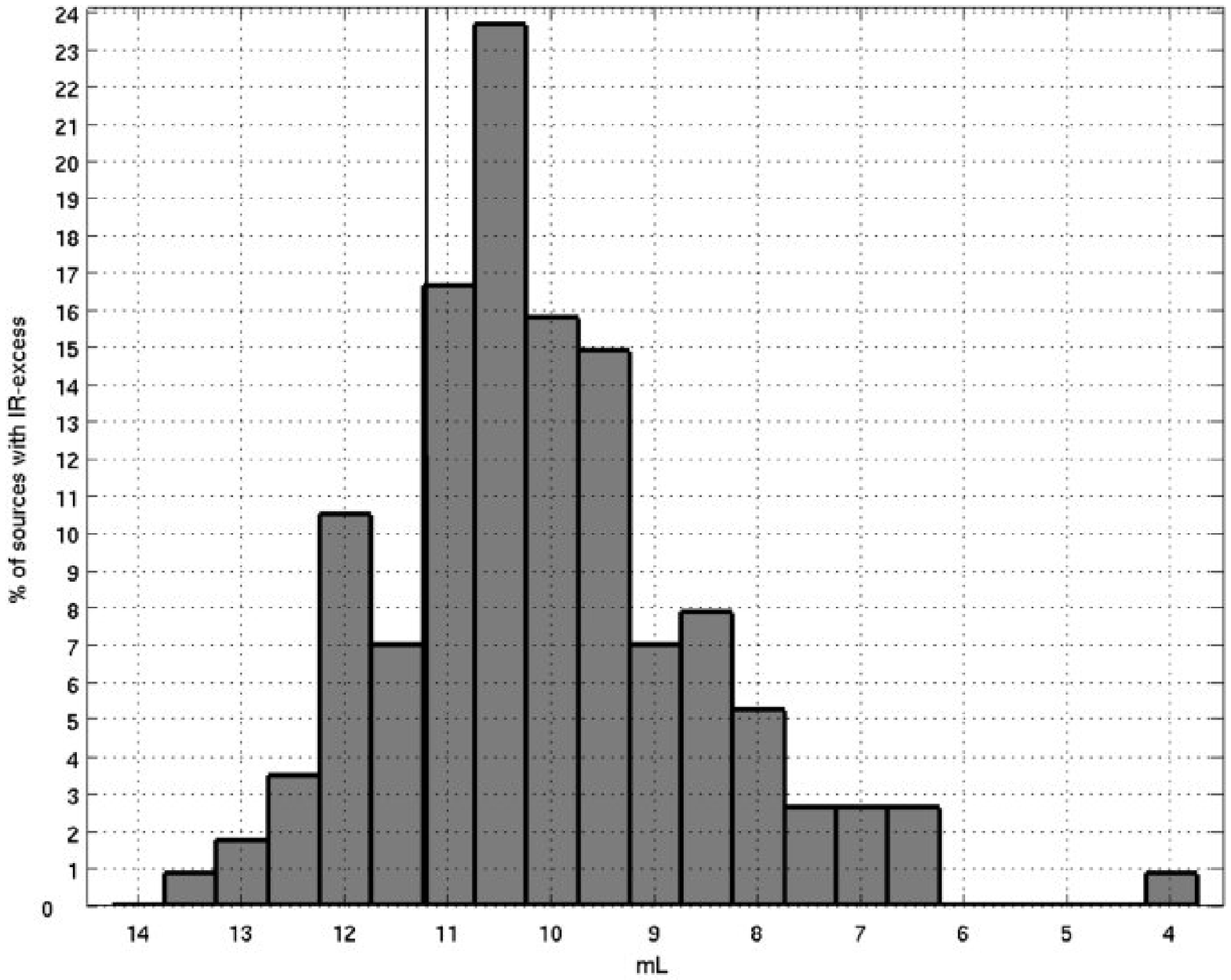}
   \caption{The percentage of all sources with an IR-excess within each magnitude interval. 
The width of the distribution is narrower than in Figure~\ref{luma}, but covers the
same range of spectral types and masses. The vertical line shows the 90\% completeness limit.}

              \label{lumir}
    \end{figure}


\section{Discussion}
\label{discussion}
\subsection{IR-excess as an indicator of circumstellar disks}
\label{IRind}

The location of stars in the JHK$_s$L diagram (Fig.~\ref{jhkl}) shows that $55\pm2\%$ of the sources lie outside
of the reddening vector defined by interstellar extinction. These sources have an IR-excess, displaying more
emission than simply a reddened stellar photosphere. As discussed in \S \ref{intro}, the position of these
stars in the colour-colour diagram can be explained by models of circumstellar disks around classical
T Tauri stars and AeBe stars, depending on the characteristics of the disk (e.g. whether these have central 
holes, varying inclination angles etc) (Lada \& Adams~\cite{ladaadams}). The IR-excess stars here are therefore 
interpreted as stars with circumstellar disks, with the fraction of IR-excess stars taken to be equal to the 
cluster disk fraction (CDF). The presence of disks is also supported by the presence of the CO 2.3 $\mu$m 
bandhead emission or absorption, found in four objects in RCW 57 (Figuer\^edo et al.~\cite{figueredoetal}). 
The preferred model for such molecular emission is that it arises from the inner portions of disks
(Barbosa et al.~\cite{barbosaetal}). Such an identification has been made, for example,  for CO bandhead emission seen in the massive star formation region of M17, where optical spectra of some of the stars also indicate the presence of 
circumstellar disks (Hanson et al.~\cite{hansonetal}; Barbosa et al.~\cite{barbosaetal}).

\subsection{Spatial Distribution of IR-excess sources}
\label{distribution}

\begin{figure}
  \centering
  \includegraphics[width=8.8cm]{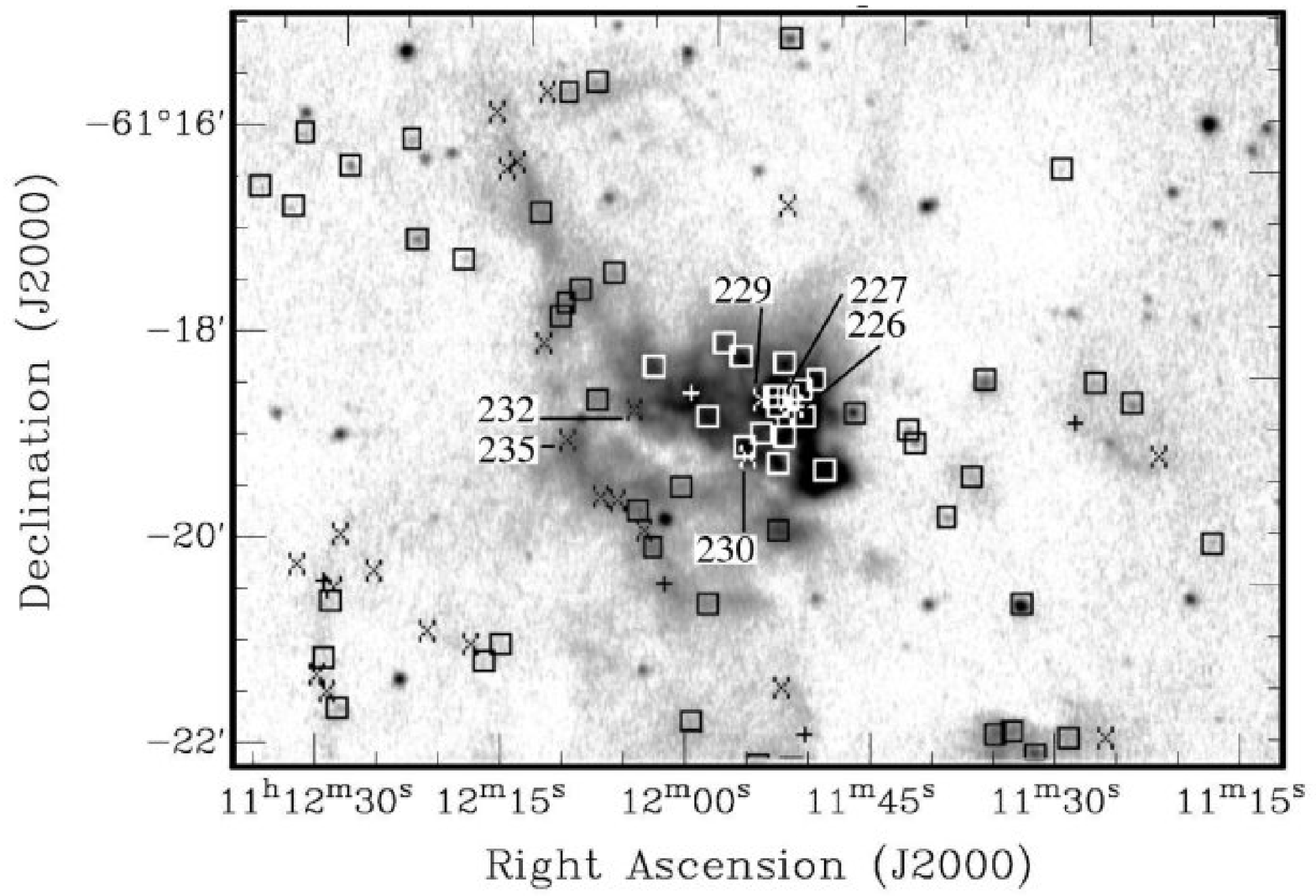}
  \caption{Spatial distribution of IR-excess sources in RCW 57. The underlying image is the SPIREX L-band image. Squares mark stars with an IR-excess 
found in the JHK$_s$L-bands, plus signs mark stars with an IR-excess found only in the K$_s$L-bands
and crosses mark stars with an IR-excess only found in L-band (there is no difference between white/black
symbols, with the colour chosen in the nebulous regions for clarity). The stars with labels are the six
reddest sources with (K$_s$-L)$>$5 and are only seen in L-band.}
              \label{spafig}
    \end{figure}

The spatial distribution of IR-excess stars can be seen in Figure~\ref{spafig}. Squares represent sources
detected in all four bands, plus signs sources only found in K$_s$-and L-band and crosses indicate the positions
of sources only found in L-band. The underlying image is the SPIREX L-band image. Sources with IR-excess are mainly located in the central cluster and follow 
the nebulous arcs outside the cluster. Stars found in all four bands dominate the cluster, while stars only 
found in L-band predominantely lie in the nebulous arcs away from the centre. However, four of the six
reddest stars lie in the central cluster (labelled \#220, \#227, \#229 and \#230 in Fig.~\ref{spafig})
(\S~\ref{highmass}). Table~\ref{region} shows how
the CDF varies at increasing radii from the centre, and Table~\ref{radial} shows how the CDF and surface 
density of all detected stars and IR-excess stars varies in regions at increasing distances from the centre. 
There is a slight increase in disk fraction and density at a distance of 3$\arcmin$-4$\arcmin$ which coincides 
with the nebulous arcs to the northeast and east of the central cluster. Figuer\^edo et al. 
(\cite{figueredoetal}) find a gradient in the spatial distribution of near IR (H-K) colour indices towards the southwest.
They suggest that this indicates a progression of star formation from the northeast to the southwest.
Fig.~\ref{spafig} however shows IR-excess sources throughout the source, favoring star formation without any
particular preferred gradient.

\begin{table}
\caption{Cluster Disk Fraction as a function of angular distance from the centre of RCW 57 (central coordinates are RA 11h11m53.41s, Dec -61$\degr$18$\arcmin$22.5$\arcsec$) based on the fraction of stars with an IR excess within each radius.}
\label{region}
\centering
\begin{tabular}{c c}
\hline\hline
radius & CDF \\
(arcmin) & \% \\
\hline
$<$1 & 95$\pm$1\\
$<$2 & 90$\pm$1\\
$<$3 & 79$\pm$2\\
$<$4 & 73$\pm$2\\
$<$5 & 66$\pm$1\\
$<$6 & 62$\pm$2\\
$<$7 & 58$\pm$2\\
$>$7 & 55$\pm$2\\
\hline
\end{tabular}
\end{table}

\begin{table}
\caption{Variation of the cluster disk fraction and the surface density of sources in annuli at increasing 
angular distance from the centre (same central coordinates as in Table~\ref{region}). The disk fraction 
decreases with increasing angular distance, though rises slightly however at a distance of 3-4 arcmin. The
surface density of all stars and IR-excess stars behaves similarly. The slight increase coincides with 
the nebulous arcs to the northeast and east of the cluster seen in Figure~\ref{spafig}, where many of the 
sources only detected in L-band lie. Errors in the values for the surface density are $\pm 0.1$ sources
per arcmin$^2$.}
\label{radial}
\centering
\begin{tabular}{c c c c}
\hline\hline
distance & fraction & no. stars & no. IR-excess sources \\
(arcmin) & \% & (per arcmin$^2$) & (per arcmin$^2$) \\
\hline
0-1 & 95 $\pm$ 1 & 7 & 6.7\\
1-2 & 82 $\pm$ 1 & 1.8 & 1.5\\
2-3 & 59 $\pm$ 5 & 1.4 & 0.8\\
3-4 & 64 $\pm$ 3 & 1.6 & 1.0\\
4-5 & 50 $\pm$ 2 & 1.6 & 0.8\\
5-6 & 45 $\pm$ 7 & 0.8 & 0.4\\
6-7 & 27 $\pm$ 5 & 0.5 & 0.1\\
\hline
\end{tabular}
\end{table}

\subsection{The Cluster Disk Fraction}
\label{diskevo}

The cluster disk fraction determined here, based on the colour-colour and colour-magnitude diagrams
(Figs.~\ref{jhkl} to~\ref{cmd}) for the entire region of the image, is $\sim$54\% when excluding potential foreground stars (Table~\ref{IRexcess}). However, as discussed earlier this is probably a lower limit due to
the contamination of background sources in the outer regions of the image. The CDF together with the ages of 
the cluster, the CDF can give information about the evolution and lifetime of circumstellar disks. In particular, since RCW 57 is a region of massive star formation (see \S~\ref{rcw57} and \S~\ref{mass}), it is possible to gain valuable information about the formation of massive stars and their circumstellar disks. Unfortunately, the age of RCW 57 is not well known. However, there are several indicators for recent and ongoing star formation, such as CO 2.3 $\mu$m bandhead emission and absorption, the detection of CH$_3$OH and H$_2$O maser sources and the clustering of IR-excess sources in the central region (\S~\ref{rcw57} 
and \S~\ref{distribution}). Within a radius of $\sim$4\arcmin ~of the centre of the cluster, the fraction of sources with an IR-excess is $\sim$73\% (Table~\ref{region}). This high IR-excess fraction, interpreted as the cluster disk fraction (CDF) (\S~\ref{IRind}), suggests that the region is young. These signs of early stages 
of star formation, combined with the fast evolution timescales for massive pre-main sequence stars (on 
the order of 10$^5$ to 10$^6$ years), indicates an upper limit for the age of RCW 57 of $\sim$1-2 Myr, and an
age spread within this limit.

\begin{figure}
   \centering
   \includegraphics[width=8.8cm]{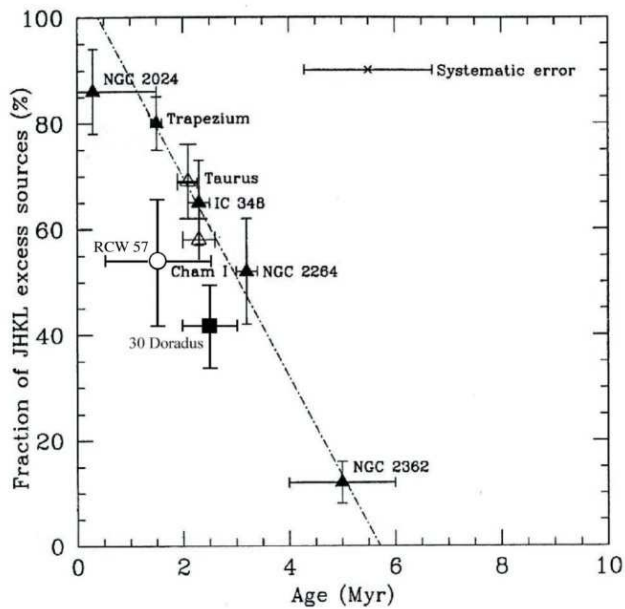}
   \caption{Cluster disk fraction (CDF) for 7 clusters by Haisch et al. (\cite{haischetal}) and 
30 Doradus (Paper I) vs their mean ages. The CDFs were determined using JHK$_s$L data as described in the text
(\S~\ref{diskevo}) and Paper I. The location of RCW 57 is indicated by a circle with error bars indicating 
the uncertainty in age and disk fraction. The dot-dashed line is the best fit determined by Haisch et al. 
(\cite{haischetal}).}
              \label{age}
    \end{figure}

In Fig.~\ref{age}, the resulting CDF and age of RCW 57 are compared to earlier JHKL surveys of clusters
in the Galaxy (Haisch et al.~\cite{haischetal}) and the results for 30 Doradus from Paper I. For the
purpose of comparison we have assumed an age for RCW 57 of 1.5$\pm$1 Myrs.  The data 
from Haisch et al. was determined in a similar way to this paper using JHKL colour-colour diagrams. The ages 
for the Trapezium, IC 348, NGC 2264 and NGC 2024 were determined using pre-main-sequence (PMS) tracks, the 
ages for NGC 2362 and NGC 1960 were determined using post-main-sequence isochrone fitting in HR-diagrams
(Haisch et al.~\cite{haischetal}). The systemetic error in the upper right corner of Fig.~\ref{age}
gives their estimate of the overall systematic error introduced in using different PMS tracks. The 
clusters included cover a range of masses and ages (0.3 Myr to 30 Myr). The error in the CDF for RCW 57 
is $\pm$12\% and includes the error due to the uncertainty in the number of IR-excess sources ($\pm$2\%).
It also allows for the uncertainty in determining the number of foreground stars. The position of RCW 57 
in Fig.~\ref{age} lies below the least-squares straight line fit to the data of Haisch et al.. However, this 
position on the diagram is clearly uncertain. The CDF determined includes data covering the entire SPIREX 
image, although the outer regions might be biased towards stars not belonging to RCW 57. If we only take
the inner 5\arcmin-6\arcmin into account (cf. the region considered by Persi et al.~\cite{persietal}),
this increases the CDF to 60-65\%. In the central arcmin the disk fraction is even higher, $\sim$95\%.
Considering  the additional uncertainty in the age, our data is consistent
with the predictions of Haisch et al. for the relation between disk fraction and age. Certainly, the disk
fraction is much higher than that determined for the oldest cluster NGC 2362. However, as with 30 Doradus
(Paper I), our best estimates of the disk fraction as a function of age lie below the relation determined
by Haisch et al. Both 30 Doradus and RCW 57 are high mass star formation regions, where the evolution 
timescales for the most massive members is expected to be less than 10$^{6}$ yrs. Nevertheless, if the CDF
had been determined from the JHK$_s$ data alone, the result would have appeared to fall significantly below
that of the Haisch et al. data. In RCW 57, it is possible that the external photoevaporation of 
circumstellar disks decreases the CDF. Lifetimes for photoevaporated disks may then be of order of 0.1 Myr 
(cf. Hollenbach et al.~\cite{hollenbachetal}). If this is true for RCW 57, the initial disk fraction would 
be higher than determined here from the JHK$_s$L data. Extrapolating the best fit suggests that the age for 
circumstellar disks is no more than $\sim$6 Myr.

\subsection{The Cluster Core}
\label{highmass}

Several very red sources ((K$_s$-L)$>$4.5) lie in the central region (r $<$ 1\arcmin) and in the bright arc
to the southeast of the central cluster (Fig.~\ref{spafig}). The six reddest sources ((K$_s$-L)$>$5) in our 
data were only detected in the L-band, with their colours therefore only being lower limits. These are 
labelled in Fig.~\ref{spafig}. Four of these lie in the central cluster close to the position of IRS 1
(\#226, \#227, \#229 and \#230). The remaining two lie in the bright arc to the southeast (\#232 and \#235). 
While we cannot classify these sources based on a single measurement, Barbosa et al. (\cite{barbosaetal})
examined the cluster core in the mid-IR, determining spectral indices for 3 sources.
Comparing these to indices for  low mass protostars, they find two (Nos. 48, 50) that would be classified 
as Class I objects (two of the four sources in IRS 1 that were resolved by Barbosa et al., see 
\S~\ref{rcw57}). Silicate absorption features in the spectra of the five objects seen at 10$\mu$m by 
Frogel \& Persson (\cite{frogelpersson}) in the inner 1.4 square arc minutes, also 
indicates Class I protostellar objects. Barbosa et al. (\cite{barbosaetal}) overlay a K-band image of the
cluster core with the radio continuum map by DePree et al. (\cite{depreeetal}). None of their sources can
be associated with the radio peak emission and they suggest that the ionizing source lies hidden behind
dark clouds in the line of sight. The position of the radio peak emission is confirmed by Walsh et al.
(\cite{walshetal}). As with the results from Barbosa et al., none of our L-band sources can be connected
with the radio continuum peak, which coincides with the dark lane just south of the bright cluster core
(RA 11h11m51s, Dec -61\degr18\arcmin45\arcsec ~(J2000)). The sources we see only at L-band are likely to be 
massive Class I protostars with circumstellar disks, still heavily embedded in their dust shells (Lada et 
al.~\cite{ladaetal}). The fact that they are not detected at shorter wavelengths, and occupy the same 
region in the colour-magnitude diagram as the IR-excess stars, gives further support to this suggestion.
In addition, 22 sources only seen in the L-band are found in the nebulous arcs further away from the 
cluster core (see Fig.~\ref{spafig}). This indicates the possible presence of massive protostars in these 
regions as well, suggesting that massive star formation is not just confined to the central cluster.

\section{Conclusions}

As the second of two papers on IR-excess in massive star forming regions measured by the SPIREX telescope
at the South Pole, L-band photometry for the giant HII region RCW 57 has been presented. The L-band 
photometry from SPIREX was combined with JHK$_s$ data from 2MASS to determine the fraction of sources with 
IR-excess in JHK$_s$L colour-colour and colour magnitude diagrams. As for 30 Doradus, it is apparent that the 
JHK$_s$ data alone would considerably underestimate the fraction of IR-excess sources, with only 25 sources
classified as having an IR-excess. Using the JHK$_s$L data, 75 are counted towards the total fraction of 
stars with an IR-excess in the JHK$_s$L diagram. More than 50\% of detected L-band sources have an IR-excess. 
This is, however, likely to still be a lower estimate, since foreground contamination has not fully been taken 
into account. Limiting the analysis to the inner $\sim$6\arcmin~of the source, the disk fraction increases to 
$\sim$65\%. The results were compared to earlier surveys (Haisch et al.~\cite{haischetal}) of clusters of 
varying ages and masses in the Galaxy (ages between 0.3 Myr for NGC 2024 to 4.5 Myr for NGC 2362; masses down to 0.13 M$_{\sun}$ for NGC 2024). Although the CDF for RCW 57 lies at the lower end of what is predicted
by Haisch et al., it is still consistent with their data, confirming a very high initial disk fraction 
($>$80\%) and a lifetime of $\lesssim$6 Myr. However, our results for both RCW 57 and 30 Doradus (Paper I) suggest a faster evolution of circumstellar disks around high mass stars, since the disk fractions appear to be slightly lower. This could be caused by photoevaporation of the disks due to the intense radiation environment generated by high mass stars.

\begin{acknowledgements}
This work could not have been conducted without the great help
received from many colleagues within the US CARA and the Australian
JACARA organisations whose efforts made the SPIREX/Abu project at the
South Pole such a success, to whom we are extremely grateful.  We also
acknowledge the funding support from the Australian Research Council
and the Australian Major National Research Facilities program that
made this work possible in the first place. 
We thank the referee, A. P. Marston, for his perceptive comments, which have 
greatly improved this paper. 
This publication makes use of data products from the Two Micron 
All Sky Survey, which is a joint project of the University of Massachusetts and the Infrared 
Processing and Analysis Center/California Institute of Technology, funded by the National 
Aeronautics and Space Administration and the National Science Foundation. 
This research has made use of the NASA/ IPAC Infrared Science Archive, which is operated by 
the Jet Propulsion Laboratory, California Institute of Technology, under contract with the 
National Aeronautics and Space Administration.
\end{acknowledgements}

\begin{longtable}{c c c c c c c c c c c c}
\caption{\label{photres}(The complete table (251 rows) is only available in electronic form at the CDS). 
Magnitudes for all sources (including foreground sources and sources below the 90\% completeness limit) 
in RCW 57. Stars with measurements in all four bands are listed first. Then stars with measurements 
in K$_s$ and L only, and finally stars just detected in L-band. Column 1 gives the source id. Columns 2 and 3 
the RA and Dec respectively in J2000. The coordinates for sources found in all four bands are from the 
2MASS point source catalogue (PSC), positions for the remaining stars are determined by reference to 2MASS 
images. Columns 4, 6, 8 and 10 give the JHK$_s$- and L-band magnitudes respectively. Columns 5, 7, 9 and
11 give the photometric errors. For sources detected in all bands, the JHK$_s$ magnitudes and errors
are taken from the 2MASS PSC. A `null' as error indicates a 95\% confidence upper limit for the 2MASS
magnitude in the PSC. The L-band errors are combined from the errors in daophot and the errors due to the
zero point correction. Sources only detected in the K$_s$- and L-bands have magnitudes determined from this work. 
For sources not detected at J, H or K$_s$ the upper limits on these magnitudes are 15.8, 15.1 and 14.3 
respectively. Stars with an IR-excess are marked with an `e' in Col. 12 (comments). 
Likely foreground stars are marked with `fg'. Stars that match sources in Persi et al. 
(\cite{persietal}) are marked with `Persi' together with the id assigned in their paper. Stars 
that match the MIR sources in Barbosa et al. (\cite{barbosaetal}) are marked with `No' and the id 
used in their paper.}\\
\hline\hline
id & RA (J2000) & Dec (J2000) & $m_{J}$ & $\sigma_{J}$ & $m_{H}$ & $\sigma_{H}$ & $m_{K}$ & $\sigma_{K}$ & $m_{L}$ & $\sigma_{L}$ & comments\\
& (h m s) & (d m s)& & & & & & & & &\\
\hline
\endfirsthead
\caption{continued.}\\
\hline
id & RA (J2000) & Dec (J2000) & $m_{J}$ & $\sigma_{J}$ & $m_{H}$ & $\sigma_{H}$ & $m_{K}$ & $\sigma_{K}$ & $m_{L}$ & $\sigma_{L}$ & comments\\
& (h m s) & (d m s)& & & & & & & & &\\
\hline
\endhead
\hline
\endfoot
1   & 11 10 50.0 & -61 18 20.2 & 09.8 & 0.02 & 08.9 & 0.03 & 08.6 & 0.02 & 08.3 & 0.06 & \\
2   & 11 10 56.8 & -61 17 08.6 & 09.7 & 0.02 & 08.9 & 0.05 & 08.5 & 0.02 & 08.5 & 0.04 & \\
3   & 11 10 59.1 & -61 17 58.3 & 12.6 & 0.02 & 10.7 & 0.02 & 09.9 & 0.02 & 09.3 & 0.07 & \\
4   & 11 10 59.9 & -61 18 39.8 & 15.1 & 0.06 & 14.6 & 0.12 & 14.3 & 0.08 & 09.7 & 0.04 & e\\
5   & 11 11 05.2 & -61 15 32.8 & 10.4 & 0.02 & 09.7 & 0.02 & 09.4 & 0.02 & 09.4 & 0.06 & \\
6   & 11 11 06.0 & -61 15 05.3 & 11.5 & 0.02 & 10.6 & 0.02 & 10.3 & 0.02 & 10.0 & 0.12 & \\
7   & 11 11 09.6 & -61 15 36.1 & 11.0 & 0.02 & 10.3 & 0.02 & 10.0 & 0.02 & 10.2 & 0.05 & \\
8   & 11 11 10.6 & -61 17 45.2 & 08.2 & 0.02 & 07.8 & 0.05 & 07.6 & 0.02 & 07.5 & 0.03 & fg\\
9   & 11 11 12.4 & -61 15 20.4 & 11.4 & 0.02 & 10.8 & 0.02 & 10.7 & 0.02 & 11.4 & 0.04 & \\
10  & 11 11 12.4 & -61 19 08.9 & 12.1 & 0.02 & 10.4 & 0.02 & 09.7 & 0.02 & 10.1 & 0.05 & \\
11  & 11 11 12.7 & -61 17 04.9 & 10.6 & 0.02 & 09.6 & 0.02 & 09.3 & 0.02 & 09.0 & 0.05 & \\
12  & 11 11 15.0 & -61 15 26.5 & 10.5 & 0.02 & 09.7 & 0.02 & 09.2 & 0.02 & 08.9 & 0.04 & \\
13  & 11 11 16.7 & -61 21 19.7 & 18.1 & null & 13.9 & 0.02 & 12.0 & 0.02 & 10.6 & 0.04 & \\
14  & 11 11 16.7 & -61 12 51.1 & 08.7 & 0.01 & 07.6 & 0.03 & 07.2 & 0.01 & 06.8 & 0.03 & \\
15  & 11 11 17.7 & -61 19 39.7 & 12.2 & 0.02 & 12.1 & 0.05 & 12.0 & 0.02 & 10.6 & 0.03 & e, fg\\
16  & 11 11 17.8 & -61 18 00.0 & 10.9 & 0.02 & 10.8 & 0.02 & 10.7 & 0.02 & 11.6 & 0.04 & \\
17  & 11 11 18.6 & -61 16 25.6 & 11.1 & 0.02 & 10.2 & 0.02 & 09.8 & 0.02 & 09.5 & 0.05 & \\
18  & 11 11 19.5 & -61 20 09.5 & 12.4 & 0.03 & 10.2 & 0.02 & 09.2 & 0.02 & 08.6 & 0.03 & \\
19  & 11 11 19.7 & -61 15 26.0 & 08.0 & 0.01 & 07.4 & 0.04 & 07.3 & 0.02 & 07.2 & 0.03 & \\
20  & 11 11 20.1 & -61 17 24.0 & 12.2 & 0.04 & 11.1 & 0.04 & 10.7 & 0.03 & 10.2 & 0.09 & \\
21  & 11 11 21.2 & -61 18 43.0 & 14.8 & 0.05 & 13.5 & 0.05 & 12.9 & 0.05 & 12.0 & 0.35 & \\
22  & 11 11 22.4 & -61 16 07.8 & 10.4 & 0.02 & 09.6 & 0.02 & 09.3 & 0.02 & 09.0 & 0.04 & \\
23  & 11 11 24.0 & -61 17 22.7 & 11.1 & 0.02 & 10.4 & 0.02 & 10.2 & 0.02 & 10.1 & 0.08 & \\
24  & 11 11 24.1 & -61 17 12.3 & 12.3 & 0.02 & 11.5 & 0.02 & 10.9 & 0.02 & 10.7 & 0.04 & \\
25  & 11 11 24.4 & -61 20 49.5 & 17.4 & null & 13.6 & 0.03 & 11.8 & 0.02 & 10.5 & 0.04 & \\
26  & 11 11 24.9 & -61 18 16.6 & 10.5 & 0.02 & 10.3 & 0.02 & 10.2 & 0.02 & 09.9 & 0.05 & e, fg\\
27  & 11 11 26.5 & -61 13 06.6 & 12.8 & 0.03 & 11.1 & 0.02 & 10.5 & 0.02 & 09.9 & 0.09 & \\
28  & 11 11 26.6 & -61 18 39.8 & 14.4 & 0.04 & 12.0 & 0.03 & 10.9 & 0.03 & 09.7 & 0.04 & \\
29  & 11 11 27.1 & -61 17 03.9 & 15.0 & 0.05 & 13.7 & 0.06 & 13.1 & 0.05 & 11.7 & 0.09 & \\
30  & 11 11 27.7 & -61 16 12.7 & 13.9 & 0.03 & 13.6 & 0.08 & 13.5 & 0.07 & 12.1 & 0.11 & \\
31  & 11 11 27.9 & -61 17 01.4 & 16.4 & 0.15 & 15.7 & 0.14 & 14.8 & 0.14 & 11.7 & 0.05 & \\
32  & 11 11 28.4 & -61 18 08.6 & 15.3 & 0.06 & 14.3 & 0.05 & 13.6 & 0.08 & 10.5 & 0.08 & e\\
33  & 11 11 29.3 & -61 21 36.1 & 14.8 & null & 14.8 & 0.15 & 14.1 & 0.17 & 10.1 & 0.07 & e, fg\\
34  & 11 11 30.1 & -61 17 23.6 & 13.0 & 0.03 & 11.0 & 0.02 & 10.2 & 0.02 & 09.6 & 0.06 & \\
35  & 11 11 31.4 & -61 16 04.6 & 15.1 & 0.06 & 14.9 & 0.09 & 14.6 & 0.10 & 10.6 & 0.04 & e, fg\\
36  & 11 11 31.6 & -61 17 23.7 & 13.2 & 0.03 & 11.8 & 0.02 & 11.1 & 0.02 & 10.4 & 0.04 & Persi 4\\
37  & 11 11 31.6 & -61 22 05.1 & 15.7 & 0.15 & 14.5 & 0.15 & 13.7 & 0.11 & 10.6 & 0.09 & e\\
38  & 11 11 31.7 & -61 21 46.9 & 10.4 & 0.03 & 09.8 & 0.03 & 09.4 & 0.03 & 08.8 & 0.04 & e\\
39  & 11 11 31.9 & -61 16 12.6 & 14.9 & 0.04 & 14.3 & 0.04 & 13.8 & 0.06 & 11.3 & 0.06 & \\
40  & 11 11 32.0 & -61 18 06.4 & 14.9 & 0.04 & 14.5 & 0.04 & 14.4 & 0.10 & 12.2 & 0.06 & \\
41  & 11 11 33.0 & -61 23 51.8 & 11.7 & 0.02 & 09.2 & 0.02 & 08.1 & 0.03 & 07.3 & 0.03 & \\
42  & 11 11 33.4 & -61 20 19.7 & 16.8 & null & 15.7 & 0.22 & 10.4 & 0.02 & 07.5 & 0.03 & e\\
43  & 11 11 33.6 & -61 21 35.5 & 14.1 & null & 13.2 & null & 14.0 & 0.21 & 09.2 & 0.05 & e, fg\\
44  & 11 11 33.9 & -61 11 35.0 & 09.0 & 0.03 & 08.2 & 0.04 & 07.8 & 0.02 & 07.5 & 0.03 & \\
45  & 11 11 34.3 & -61 13 16.8 & 12.3 & 0.02 & 10.9 & 0.02 & 10.3 & 0.02 & 09.7 & 0.07 & \\
46  & 11 11 34.6 & -61 21 36.9 & 13.9 & 0.08 & 12.6 & 0.10 & 11.8 & 0.07 & 08.7 & 0.06 & e\\
47  & 11 11 35.3 & -61 18 25.3 & 15.8 & 0.09 & 14.3 & 0.07 & 13.4 & 0.04 & 12.0 & 0.08 & \\
48  & 11 11 36.4 & -61 16 55.0 & 16.3 & 0.13 & 15.2 & 0.09 & 14.6 & 0.09 & 11.9 & 0.08 & \\
49  & 11 11 36.6 & -61 12 04.6 & 15.8 & 0.08 & 15.2 & 0.09 & 15.0 & 0.14 & 10.8 & 0.05 & e\\
50  & 11 11 37.1 & -61 18 08.6 & 13.6 & 0.04 & 12.0 & null & 11.1 & null & 08.7 & 0.05 & e\\
51  & 11 11 38.0 & -61 12 05.2 & 12.7 & 0.03 & 11.5 & 0.02 & 11.2 & 0.02 & 10.2 & 0.03 & e\\
52  & 11 11 38.1 & -61 19 04.3 & 12.6 & 0.03 & 12.1 & 0.04 & 11.9 & 0.04 & 09.9 & 0.06 & e\\
53  & 11 11 38.6 & -61 22 22.4 & 16.8 & null & 17.0 & null & 12.1 & 0.02 & 08.6 & 0.04 & e\\
54  & 11 11 39.5 & -61 11 34.9 & 15.6 & 0.08 & 15.0 & 0.08 & 15.1 & 0.17 & 11.0 & 0.04 & e, fg\\
55  & 11 11 39.6 & -61 13 54.6 & 12.3 & 0.04 & 11.5 & null & 11.3 & 0.03 & 11.7 & 0.04 & \\
56  & 11 11 39.7 & -61 19 30.1 & 16.3 & null & 14.5 & null & 13.3 & 0.08 & 10.6 & 0.04 & e\\
57  & 11 11 39.8 & -61 14 49.2 & 13.7 & 0.02 & 11.6 & 0.02 & 10.7 & 0.02 & 10.3 & 0.05 & \\
58  & 11 11 41.0 & -61 20 22.1 & 09.7 & 0.02 & 09.1 & 0.02 & 08.9 & 0.02 & 08.9 & 0.03 & \\
59  & 11 11 41.1 & -61 22 29.1 & 17.6 & null & 13.4 & 0.05 & 10.3 & 0.02 & 08.2 & 0.03 & \\
60  & 11 11 42.5 & -61 16 25.2 & 11.6 & 0.02 & 09.7 & 0.03 & 08.9 & 0.02 & 08.2 & 0.03 & \\
61  & 11 11 42.5 & -61 18 15.6 & 17.9 & null & 14.6 & 0.06 & 12.7 & 0.03 & 11.0 & 0.04 & \\
62  & 11 11 42.6 & -61 18 46.5 & 17.0 & null & 16.8 & null & 12.7 & 0.02 & 10.7 & 0.04 & e\\
63  & 11 11 42.9 & -61 12 57.2 & 11.8 & 0.02 & 11.0 & 0.02 & 10.6 & 0.02 & 09.9 & 0.08 & e\\
64  & 11 11 43.2 & -61 18 39.2 & 16.1 & null & 14.5 & null & 13.5 & 0.07 & 10.1 & 0.05 & e\\
65  & 11 11 43.2 & -61 19 15.6 & 16.4 & null & 15.0 & 0.10 & 13.9 & 0.06 & 12.9 & 0.15 & \\
66  & 11 11 43.5 & -61 14 33.5 & 12.7 & 0.04 & 10.6 & 0.03 & 09.6 & 0.02 & 08.9 & 0.04 & \\
67  & 11 11 45.1 & -61 14 24.3 & 16.5 & 0.15 & 15.0 & 0.09 & 14.3 & 0.09 & 11.0 & 0.04 & e\\
68  & 11 11 45.4 & -61 21 18.2 & 14.6 & 0.03 & 14.2 & 0.03 & 14.1 & 0.06 & 12.2 & 0.13 & \\
69  & 11 11 46.2 & -61 14 30.7 & 15.8 & 0.08 & 14.2 & 0.04 & 13.6 & 0.06 & 11.7 & 0.04 & \\
70  & 11 11 46.9 & -61 11 31.9 & 09.3 & 0.02 & 09.1 & 0.02 & 09.1 & 0.02 & 09.0 & 0.07 & fg\\
71  & 11 11 47.1 & -61 13 49.0 & 10.3 & 0.02 & 09.7 & 0.02 & 09.5 & 0.02 & 10.6 & 0.05 & \\
72  & 11 11 47.3 & -61 15 25.0 & 13.5 & 0.02 & 11.7 & 0.02 & 10.9 & 0.02 & 10.5 & 0.04 & \\
73  & 11 11 47.7 & -61 20 44.5 & 16.1 & 0.10 & 15.8 & 0.14 & 15.1 & null & 11.8 & 0.06 & \\
74  & 11 11 47.7 & -61 18 30.6 & 12.5 & 0.03 & 10.9 & 0.02 & 10.0 & 0.02 & 08.4 & 0.03 & e, Persi 43, No 04\\
75  & 11 11 47.8 & -61 16 17.3 & 13.4 & 0.04 & 11.6 & 0.03 & 10.7 & 0.02 & 09.7 & 0.05 & \\
76  & 11 11 47.9 & -61 13 24.7 & 10.2 & 0.02 & 09.7 & 0.02 & 09.5 & 0.02 & 09.8 & 0.05 & fg\\
77  & 11 11 49.4 & -61 13 09.3 & 10.3 & 0.02 & 08.9 & 0.07 & 08.4 & 0.02 & 08.0 & 0.03 & \\
78  & 11 11 49.6 & -61 19 08.5 & 14.4 & null & 13.3 & 0.29 & 10.7 & 0.15 & 06.7 & 0.06 & e, Persi 45, No 11\\
79  & 11 11 50.1 & -61 16 30.0 & 17.4 & null & 14.7 & 0.05 & 12.7 & 0.02 & 13.5 & 0.19 & \\
80  & 11 11 50.2 & -61 20 22.6 & 10.6 & 0.02 & 09.9 & 0.02 & 09.8 & 0.02 & 09.6 & 0.05 & \\
81  & 11 11 50.2 & -61 12 58.6 & 12.3 & 0.02 & 10.9 & 0.02 & 10.4 & 0.02 & 10.0 & 0.04 & \\
82  & 11 11 51.1 & -61 18 14.3 & 12.6 & null & 12.8 & 0.34 & 11.4 & 0.27 & 07.6 & 0.06 & e\\
83  & 11 11 51.2 & -61 14 53.5 & 12.8 & 0.02 & 11.2 & 0.02 & 10.5 & 0.02 & 10.8 & 0.04 & \\
84  & 11 11 51.7 & -61 22 04.0 & 16.2 & null & 15.0 & null & 15.0 & 0.17 & 10.9 & 0.11 & e\\
85  & 11 11 51.8 & -61 18 20.4 & 13.0 & 0.16 & 12.2 & 0.03 & 09.4 & null & 07.1 & 0.06 & e\\
86  & 11 11 51.8 & -61 14 20.5 & 13.0 & 0.02 & 11.4 & 0.02 & 10.8 & 0.02 & 10.3 & 0.12 & \\
87  & 11 11 52.0 & -61 18 37.4 & 13.6 & 0.10 & 11.9 & 0.15 & 10.1 & 0.12 & 06.6 & 0.06 & e\\
88  & 11 11 53.2 & -61 18 22.4 & 11.2 & null & 09.9 & 0.11 & 07.5 & 0.02 & 04.1 & 0.03 & e, Persi 60, Nos. 48, 50, 60, 60b, IRS 1\\
89  & 11 11 53.3 & -61 15 05.1 & 13.3 & 0.03 & 11.8 & null & 10.8 & null & 10.1 & 0.03 & \\
90  & 11 11 53.3 & -61 19 40.2 & 14.7 & null & 12.8 & null & 12.3 & 0.16 & 08.4 & 0.06 & e\\
91  & 11 11 53.5 & -61 18 04.5 & 13.7 & null & 13.2 & 0.16 & 11.3 & 0.08 & 08.3 & 0.05 & e, Persi 62, No 52\\
92  & 11 11 53.5 & -61 18 49.5 & 16.1 & 0.13 & 15.0 & 0.15 & 11.4 & null & 08.0 & 0.03 & e, Persi 61\\
93  & 11 11 53.7 & -61 19 04.5 & 16.1 & null & 13.6 & null & 11.5 & 0.10 & 08.2 & 0.04 & e\\
94  & 11 11 53.9 & -61 18 28.3 & 11.3 & null & 10.2 & null & 10.6 & 0.07 & 07.1 & 0.05 & e\\
95  & 11 11 54.2 & -61 14 50.4 & 11.8 & 0.02 & 10.9 & 0.02 & 10.1 & 0.02 & 09.0 & 0.04 & e\\
96  & 11 11 54.2 & -61 23 29.6 & 14.9 & 0.04 & 14.3 & 0.04 & 14.1 & 0.07 & 11.2 & 0.05 & e\\
97  & 11 11 54.3 & -61 22 11.3 & 14.6 & null & 13.9 & null & 14.4 & 0.16 & 12.2 & 0.29 & \\
98  & 11 11 54.4 & -61 18 24.0 & 12.0 & 0.14 & 10.6 & 0.14 & 09.0 & 0.09 & 06.5 & 0.04 & e\\
99  & 11 11 54.4 & -61 22 04.0 & 10.5 & 0.02 & 09.7 & 0.02 & 09.3 & 0.02 & 08.7 & 0.03 & e\\
100 & 11 11 54.6 & -61 13 48.4 & 12.7 & 0.02 & 11.3 & 0.02 & 10.8 & 0.02 & 11.4 & 0.07 & \\
101 & 11 11 54.8 & -61 18 30.6 & 13.5 & 0.02 & 10.9 & null & 09.6 & null & 08.0 & 0.05 & \\
102 & 11 11 55.3 & -61 18 45.2 & 15.5 & 0.08 & 13.8 & 0.10 & 12.2 & 0.10 & 09.4 & 0.05 & e, Persi 69\\
103 & 11 11 55.4 & -61 23 09.3 & 14.7 & 0.02 & 12.5 & 0.02 & 11.5 & 0.02 & 11.0 & 0.10 & \\
104 & 11 11 55.7 & -61 14 20.0 & 16.6 & 0.17 & 15.3 & 0.10 & 15.0 & 0.15 & 10.3 & 0.05 & e\\
105 & 11 11 56.2 & -61 21 46.5 & 14.4 & 0.03 & 13.9 & 0.04 & 13.8 & 0.06 & 12.3 & 0.32 & \\
106 & 11 11 56.3 & -61 16 10.2 & 11.7 & 0.02 & 10.3 & 0.02 & 09.7 & 0.02 & 09.4 & 0.05 & \\
107 & 11 11 56.3 & -61 14 03.3 & 12.7 & 0.02 & 10.7 & 0.02 & 09.9 & 0.02 & 09.4 & 0.05 & \\
108 & 11 11 56.5 & -61 18 55.0 & 13.5 & null & 13.2 & null & 10.9 & 0.10 & 07.9 & 0.04 & e, Persi 77, No 160\\
109 & 11 11 56.6 & -61 17 18.3 & 13.5 & 0.03 & 12.5 & 0.03 & 11.7 & 0.02 & 11.1 & 0.09 & Persi 79\\
110 & 11 11 56.7 & -61 17 06.4 & 14.4 & null & 13.4 & null & 13.6 & 0.08 & 13.2 & 0.20 & \\
111 & 11 11 57.0 & -61 18 02.0 & 12.5 & 0.09 & 11.3 & 0.08 & 10.4 & 0.06 & 08.2 & 0.04 & e, No 184\\
112 & 11 11 57.8 & -61 14 30.6 & 15.0 & 0.07 & 14.5 & 0.08 & 14.2 & 0.09 & 10.3 & 0.08 & e\\
113 & 11 11 58.3 & -61 16 34.4 & 15.5 & 0.08 & 14.1 & 0.05 & 13.5 & 0.05 & 11.7 & 0.08 & \\
114 & 11 11 58.3 & -61 11 54.2 & 14.0 & 0.05 & 13.0 & 0.04 & 12.7 & 0.03 & 10.6 & 0.04 & e\\
115 & 11 11 58.7 & -61 17 51.1 & 13.4 & 0.02 & 12.1 & 0.06 & 11.4 & 0.06 & 08.7 & 0.06 & e, Persi 89\\
116 & 11 11 59.4 & -61 11 55.5 & 11.2 & 0.02 & 10.5 & 0.02 & 10.3 & 0.02 & 09.8 & 0.04 & \\
117 & 11 11 59.5 & -61 20 26.6 & 14.6 & null & 13.5 & null & 13.4 & 0.17 & 09.9 & 0.06 & e\\
118 & 11 11 59.6 & -61 18 39.8 & 12.0 & 0.09 & 11.4 & 0.14 & 10.7 & 0.12 & 08.8 & 0.05 & e, Persi 90\\
119 & 11 11 59.9 & -61 21 38.8 & 16.6 & 0.15 & 15.2 & 0.12 & 14.4 & 0.12 & 10.4 & 0.08 & e\\
120 & 11 12 00.3 & -61 15 27.1 & 14.4 & 0.05 & 12.9 & 0.05 & 12.0 & 0.03 & 11.4 & 0.04 & \\
121 & 11 12 00.6 & -61 10 50.6 & 10.6 & 0.02 & 09.3 & 0.02 & 08.9 & 0.02 & 08.4 & 0.04 & \\
122 & 11 12 01.6 & -61 19 16.7 & 16.2 & null & 15.4 & null & 14.9 & 0.20 & 10.4 & 0.06 & e\\
123 & 11 12 02.5 & -61 15 02.1 & 10.5 & 0.03 & 09.3 & 0.03 & 08.9 & 0.02 & 08.6 & 0.04 & \\
124 & 11 12 02.8 & -61 19 40.7 & 13.1 & 0.03 & 10.1 & 0.02 & 08.7 & 0.02 & 07.7 & 0.03 & Persi 106\\
125 & 11 12 03.9 & -61 19 57.9 & 13.8 & 0.04 & 13.3 & 0.07 & 13.1 & 0.08 & 09.4 & 0.06 & e, Persi 110\\
126 & 11 12 04.0 & -61 18 11.2 & 14.8 & 0.15 & 12.2 & null & 11.1 & null & 08.6 & 0.05 & e, Persi 111\\
127 & 11 12 04.1 & -61 21 11.2 & 13.0 & 0.02 & 10.8 & 0.02 & 10.0 & 0.02 & 09.4 & 0.04 & \\
128 & 11 12 04.8 & -61 22 45.2 & 16.7 & null & 15.6 & 0.12 & 14.5 & 0.08 & 12.0 & 0.07 & \\
129 & 11 12 05.2 & -61 21 47.6 & 15.0 & 0.03 & 14.6 & 0.05 & 14.7 & 0.10 & 12.3 & 0.06 & \\
130 & 11 12 05.7 & -61 19 35.0 & 15.0 & 0.05 & 14.7 & 0.04 & 14.4 & 0.11 & 09.7 & 0.05 & e, fg\\
131 & 11 12 06.1 & -61 22 31.7 & 14.0 & 0.04 & 13.7 & 0.05 & 13.7 & 0.05 & 12.3 & 0.08 & \\
132 & 11 12 06.7 & -61 22 19.3 & 14.3 & 0.03 & 12.0 & 0.02 & 11.0 & 0.02 & 10.3 & 0.07 & \\
133 & 11 12 07.2 & -61 12 28.3 & 10.5 & 0.02 & 09.7 & 0.02 & 09.5 & 0.02 & 09.4 & 0.03 & \\
134 & 11 12 07.6 & -61 17 17.3 & 11.4 & 0.03 & 10.8 & 0.03 & 10.6 & 0.04 & 10.2 & 0.06 & e, Persi 177\\
135 & 11 12 08.0 & -61 15 53.8 & 12.1 & 0.02 & 11.2 & 0.02 & 10.8 & 0.02 & 10.6 & 0.07 & \\
136 & 11 12 08.5 & -61 16 31.9 & 09.6 & 0.02 & 09.5 & 0.02 & 09.4 & 0.02 & 09.3 & 0.04 & fg\\
137 & 11 12 08.6 & -61 18 27.9 & 15.6 & null & 15.0 & null & 14.9 & 0.19 & 10.3 & 0.07 & e\\
138 & 11 12 08.8 & -61 14 23.9 & 10.8 & 0.02 & 10.6 & 0.02 & 10.5 & 0.02 & 10.3 & 0.03 & e, fg\\
139 & 11 12 09.8 & -61 15 23.6 & 17.0 & null & 14.6 & 0.09 & 12.7 & 0.05 & 10.0 & 0.06 & e\\
140 & 11 12 10.0 & -61 18 44.2 & 08.9 & 0.02 & 08.9 & 0.04 & 08.9 & 0.02 & 09.0 & 0.04 & fg, Persi123\\
141 & 11 12 10.0 & -61 25 43.6 & 10.7 & 0.02 & 09.1 & 0.02 & 08.4 & 0.02 & 08.0 & 0.04 & \\
142 & 11 12 10.2 & -61 17 26.7 & 11.2 & 0.02 & 10.6 & 0.02 & 10.5 & 0.03 & 09.4 & 0.06 & e, Persi 125\\
143 & 11 12 10.9 & -61 17 34.9 & 15.7 & 0.17 & 13.5 & null & 12.2 & null & 09.4 & 0.06 & e\\
144 & 11 12 11.2 & -61 17 41.5 & 14.6 & null & 13.6 & null & 13.7 & 0.16 & 09.4 & 0.06 & e\\
145 & 11 12 12.1 & -61 14 15.5 & 12.6 & 0.03 & 11.1 & 0.02 & 10.6 & 0.02 & 10.0 & 0.03 & \\
146 & 11 12 12.4 & -61 15 28.8 & 16.5 & 0.15 & 15.4 & null & 14.6 & null & 10.1 & 0.07 & e\\
147 & 11 12 12.8 & -61 15 13.3 & 17.5 & null & 13.7 & 0.02 & 12.1 & 0.02 & 11.7 & 0.04 & \\
148 & 11 12 13.0 & -61 14 44.6 & 12.0 & 0.02 & 11.3 & 0.02 & 11.1 & 0.02 & 13.0 & 0.25 & \\
149 & 11 12 13.3 & -61 13 38.0 & 12.5 & 0.02 & 10.9 & 0.02 & 10.2 & 0.02 & 09.8 & 0.04 & \\
150 & 11 12 13.7 & -61 22 21.4 & 14.4 & 0.02 & 12.6 & 0.02 & 11.6 & 0.02 & 10.9 & 0.04 & \\
151 & 11 12 14.3 & -61 16 43.9 & 13.2 & 0.02 & 12.9 & 0.05 & 12.8 & 0.06 & 09.5 & 0.06 & e, fg\\
152 & 11 12 14.3 & -61 12 55.7 & 13.0 & 0.04 & 11.4 & 0.03 & 10.9 & 0.02 & 09.9 & 0.03 & \\
153 & 11 12 15.4 & -61 22 50.2 & 16.2 & null & 15.5 & 0.14 & 14.8 & 0.14 & 11.8 & 0.09 & \\
154 & 11 12 15.7 & -61 21 04.0 & 15.5 & 0.08 & 14.1 & 0.04 & 13.6 & 0.06 & 10.3 & 0.03 & e\\
155 & 11 12 15.9 & -61 22 23.4 & 13.0 & 0.03 & 11.2 & 0.02 & 10.4 & 0.02 & 09.6 & 0.06 & \\
156 & 11 12 16.0 & -61 21 42.9 & 15.9 & null & 15.2 & 0.11 & 14.5 & 0.13 & 11.8 & 0.07 & \\
157 & 11 12 16.3 & -61 21 56.4 & 13.3 & 0.03 & 11.8 & 0.03 & 11.3 & 0.02 & 10.8 & 0.04 & \\
158 & 11 12 17.0 & -61 21 14.1 & 15.4 & null & 13.9 & 0.06 & 13.1 & 0.07 & 10.9 & 0.13 & e\\
159 & 11 12 17.3 & -61 13 36.3 & 12.3 & 0.02 & 11.0 & 0.02 & 10.6 & 0.02 & 09.7 & 0.04 & e\\
160 & 11 12 19.1 & -61 20 12.1 & 15.3 & 0.08 & 14.8 & 0.08 & 14.6 & 0.10 & 13.3 & 0.18 & \\
161 & 11 12 20.0 & -61 19 50.5 & 14.0 & 0.04 & 13.7 & 0.04 & 13.6 & 0.06 & 12.4 & 0.05 & \\
162 & 11 12 20.5 & -61 17 07.9 & 15.3 & 0.07 & 15.0 & 0.07 & 14.8 & 0.13 & 11.2 & 0.04 & e, fg\\
163 & 11 12 21.1 & -61 11 56.0 & 10.9 & 0.03 & 10.6 & 0.02 & 10.2 & 0.02 & 08.8 & 0.05 & e\\
164 & 11 12 21.3 & -61 16 10.5 & 10.0 & 0.02 & 09.3 & 0.02 & 09.2 & 0.02 & 09.4 & 0.05 & \\
165 & 11 12 21.6 & -61 19 48.0 & 11.1 & 0.03 & 10.6 & 0.02 & 10.5 & 0.02 & 10.4 & 0.08 & fg\\
166 & 11 12 21.7 & -61 14 35.2 & 10.4 & 0.02 & 09.1 & 0.02 & 08.6 & 0.02 & 08.2 & 0.04 & \\
167 & 11 12 23.5 & -61 16 15.2 & 13.8 & 0.03 & 11.2 & 0.03 & 10.0 & 0.02 & 09.3 & 0.05 & \\
168 & 11 12 23.9 & -61 17 03.4 & 12.9 & 0.03 & 11.8 & 0.03 & 10.8 & 0.02 & 09.4 & 0.05 & e\\
169 & 11 12 24.0 & -61 21 25.2 & 11.7 & 0.02 & 09.7 & 0.02 & 08.8 & 0.02 & 08.2 & 0.03 & \\
170 & 11 12 24.3 & -61 23 58.2 & 13.2 & 0.03 & 11.2 & 0.02 & 10.4 & 0.02 & 09.9 & 0.07 & \\
171 & 11 12 24.7 & -61 16 04.9 & 16.6 & 0.16 & 14.4 & 0.06 & 12.8 & 0.05 & 10.2 & 0.04 & e\\
172 & 11 12 24.9 & -61 19 55.7 & 10.6 & 0.02 & 10.6 & 0.02 & 10.5 & 0.02 & 11.2 & 0.04 & \\
173 & 11 12 25.4 & -61 15 11.6 & 11.2 & 0.02 & 09.1 & 0.02 & 08.2 & 0.03 & 07.5 & 0.03 & \\
174 & 11 12 25.6 & -61 14 38.0 & 12.0 & 0.04 & 10.6 & null & 10.1 & null & 09.8 & 0.03 & \\
175 & 11 12 27.0 & -61 19 55.0 & 14.2 & 0.04 & 13.0 & 0.04 & 12.2 & 0.04 & 11.6 & 0.05 & \\
176 & 11 12 28.2 & -61 21 44.9 & 15.2 & 0.25 & 13.6 & null & 12.5 & null & 09.7 & 0.03 & e\\
177 & 11 12 29.4 & -61 13 54.1 & 11.4 & 0.03 & 10.0 & 0.02 & 09.5 & 0.02 & 09.3 & 0.03 & \\
178 & 11 12 29.6 & -61 19 01.9 & 12.6 & 0.03 & 10.3 & 0.02 & 09.2 & 0.02 & 08.6 & 0.04 & \\
179 & 11 12 29.6 & -61 16 21.8 & 12.3 & 0.02 & 11.4 & 0.02 & 10.8 & 0.02 & 09.9 & 0.04 & e\\
180 & 11 12 29.6 & -61 23 27.1 & 12.4 & 0.03 & 10.4 & 0.02 & 09.5 & 0.02 & 08.8 & 0.04 & \\
181 & 11 12 29.7 & -61 24 22.1 & 13.5 & 0.02 & 11.3 & 0.02 & 10.3 & 0.02 & 09.6 & 0.08 & \\
182 & 11 12 29.7 & -61 20 42.9 & 14.3 & null & 15.3 & 0.19 & 13.4 & null & 10.3 & 0.04 & e\\
183 & 11 12 30.1 & -61 21 14.3 & 14.9 & 0.19 & 14.5 & 0.20 & 13.8 & 0.16 & 10.1 & 0.04 & e\\
184 & 11 12 31.7 & -61 23 59.1 & 12.5 & 0.02 & 10.3 & 0.02 & 09.3 & 0.02 & 08.6 & 0.04 & \\
185 & 11 12 32.9 & -61 14 13.8 & 11.9 & 0.02 & 10.7 & 0.02 & 10.2 & 0.02 & 10.3 & 0.04 & \\
186 & 11 12 33.4 & -61 15 52.1 & 11.6 & 0.02 & 10.2 & 0.02 & 09.4 & 0.02 & 08.8 & 0.04 & \\
187 & 11 12 33.4 & -61 16 05.3 & 16.2 & null & 13.6 & 0.04 & 12.0 & 0.02 & 09.9 & 0.03 & e\\
188 & 11 12 34.3 & -61 16 48.7 & 12.7 & 0.02 & 12.0 & 0.02 & 11.5 & 0.02 & 11.0 & 0.05 & e\\
189 & 11 12 34.9 & -61 18 51.5 & 14.0 & 0.02 & 11.2 & 0.02 & 09.9 & 0.02 & 08.9 & 0.06 & \\
190 & 11 12 37.0 & -61 16 39.1 & 16.0 & 0.10 & 13.9 & 0.05 & 12.7 & 0.04 & 10.5 & 0.04 & e\\
191 & 11 12 39.6 & -61 21 41.7 & 10.3 & 0.02 & 09.2 & 0.02 & 08.8 & 0.02 & 08.7 & 0.05 & \\
192 & 11 12 41.6 & -61 21 05.4 & 11.8 & 0.02 & 10.8 & 0.02 & 10.3 & 0.02 & 10.2 & 0.04 & \\
193 & 11 12 48.4 & -61 21 10.4 & 13.0 & 0.03 & 10.8 & 0.03 & 09.8 & 0.02 & 09.2 & 0.04 & \\
194 & 11 12 49.1 & -61 18 13.2 & 10.9 & 0.02 & 10.4 & 0.02 & 10.2 & 0.02 & 10.3 & 0.05 & fg\\
195 & 11 12 53.2 & -61 16 41.0 & 14.2 & 0.03 & 11.1 & 0.02 & 09.6 & 0.02 & 08.6 & 0.04 & \\
196 & 11 12 54.6 & -61 16 51.0 & 10.5 & 0.02 & 09.2 & 0.02 & 08.7 & 0.02 & 08.6 & 0.10 & \\
197 & 11 12 55.6 & -61 16 38.0 & 14.0 & 0.03 & 11.3 & 0.02 & 10.1 & 0.02 & 09.7 & 0.07 & \\
198 & 11 12 58.7 & -61 16 38.8 & 12.9 & 0.03 & 09.7 & 0.02 & 08.2 & null & 07.0 & 0.03 & \\
199 & 11 12 59.0 & -61 20 10.6 & 12.4 & 0.03 & 10.4 & 0.02 & 09.5 & 0.02 & 08.9 & 0.04 & \\
200 & 11 13 01.4 & -61 16 53.6 & 11.9 & 0.02 & 10.4 & 0.02 & 09.9 & 0.02 & 09.2 & 0.07 & \\
201 & 11 13 03.3 & -61 17 48.6 & 12.6 & 0.02 & 10.1 & 0.02 & 08.9 & 0.02 & 08.1 & 0.04 & \\
202 & 11 11 16.0 & -61 15 40.2 &  -   &  -   &  -   &  -   & 09.7 & 0.01 & 09.3 & 0.05 & \\
203 & 11 11 24.1 & -61 16 41.7 &  -   &  -   &  -   &  -   & 14.4 & 0.13 & 12.1 & 0.12 & \\
204 & 11 11 25.9 & -61 16 28.8 &  -   &  -   &  -   &  -   & 15.6 & 0.40 & 12.1 & 0.09 & \\
205 & 11 11 28.7 & -61 21 54.9 &  -   &  -   &  -   &  -   & 14.8 & 0.21 & 11.0 & 0.14 & e\\
206 & 11 11 29.3 & -61 18 33.1 &  -   &  -   &  -   &  -   & 15.6 & 0.39 & 10.8 & 0.11 & e\\
207 & 11 11 35.9 & -61 23 00.6 &  -   &  -   &  -   &  -   & 14.5 & 0.16 & 11.4 & 0.09 & \\
208 & 11 11 48.6 & -61 14 23.7 &  -   &  -   &  -   &  -   & 15.9 & 0.54 & 12.8 & 0.14 & \\
209 & 11 11 50.1 & -61 21 41.9 &  -   &  -   &  -   &  -   & 14.8 & 0.23 & 10.9 & 0.12 & e\\
210 & 11 11 51.9 & -61 14 40.7 &  -   &  -   &  -   &  -   & 14.6 & 0.17 & 13.2 & 0.09 & \\
211 & 11 12 00.8 & -61 18 26.2 &  -   &  -   &  -   &  -   & 10.8 & 0.09 & 08.1 & 0.05 & e\\
212 & 11 12 01.8 & -61 20 19.2 &  -   &  -   &  -   &  -   & 13.7 & 0.17 & 09.5 & 0.05 & e\\
213 & 11 12 08.2 & -61 14 48.7 &  -   &  -   &  -   &  -   & 09.7 & 0.02 & 09.7 & 0.03 & \\
214 & 11 12 20.0 & -61 22 40.6 &  -   &  -   &  -   &  -   & 14.9 & 0.18 & 11.5 & 0.05 & \\
215 & 11 12 20.7 & -61 20 52.4 &  -   &  -   &  -   &  -   & 15.3 & 0.27 & 11.5 & 0.06 & \\
216 & 11 12 27.3 & -61 20 13.5 &  -   &  -   &  -   &  -   & 12.7 & 0.04 & 11.5 & 0.04 & \\
217 & 11 12 29.7 & -61 20 32.1 &  -   &  -   &  -   &  -   & 13.4 & 0.08 & 10.4 & 0.04 & e\\
218 & 11 12 31.3 & -61 20 42.3 &  -   &  -   &  -   &  -   & 11.7 & 0.02 & 12.0 & 0.08 & \\
219 & 11 11 22.3 & -61 18 49.1 &  -   &  -   &  -   &  -   &  -   &  -   & 11.1 & 0.14 & e\\
220 & 11 11 25.6 & -61 21 36.9 &  -   &  -   &  -   &  -   &  -   &  -   & 10.7 & 0.12 & e\\
221 & 11 11 30.1 & -61 13 42.7 &  -   &  -   &  -   &  -   &  -   &  -   & 10.8 & 0.04 & e\\
222 & 11 11 50.1 & -61 22 38.5 &  -   &  -   &  -   &  -   &  -   &  -   & 10.4 & 0.09 & e\\
223 & 11 11 52.2 & -61 22 45.4 &  -   &  -   &  -   &  -   &  -   &  -   & 10.4 & 0.10 & e\\
224 & 11 11 52.3 & -61 22 57.0 &  -   &  -   &  -   &  -   &  -   &  -   & 10.9 & 0.15 & e\\
225 & 11 11 52.4 & -61 21 14.7 &  -   &  -   &  -   &  -   &  -   &  -   & 10.2 & 0.08 & e\\
226 & 11 11 52.4 & -61 18 28.2 &  -   &  -   &  -   &  -   &  -   &  -   & 07.2 & 0.06 & e\\
227 & 11 11 53.2 & -61 18 33.6 &  -   &  -   &  -   &  -   &  -   &  -   & 07.4 & 0.05 & e\\
228 & 11 11 53.7 & -61 16 30.6 &  -   &  -   &  -   &  -   &  -   &  -   & 10.6 & 0.04 & e\\
229 & 11 11 55.3 & -61 18 26.2 &  -   &  -   &  -   &  -   &  -   &  -   & 08.2 & 0.05 & e, Persi 68\\
230 & 11 11 56.1 & -61 19 00.5 &  -   &  -   &  -   &  -   &  -   &  -   & 08.8 & 0.05 & e, Persi 71\\
231 & 11 12 04.4 & -61 19 46.5 &  -   &  -   &  -   &  -   &  -   &  -   & 09.5 & 0.05 & e\\
232 & 11 12 05.6 & -61 18 34.3 &  -   &  -   &  -   &  -   &  -   &  -   & 08.8 & 0.06 & e\\
233 & 11 12 06.6 & -61 19 28.5 &  -   &  -   &  -   &  -   &  -   &  -   & 10.2 & 0.06 & e\\
234 & 11 12 08.1 & -61 19 26.8 &  -   &  -   &  -   &  -   &  -   &  -   & 09.7 & 0.06 & e\\
235 & 11 12 10.9 & -61 18 54.3 &  -   &  -   &  -   &  -   &  -   &  -   & 09.1 & 0.06 & e\\
236 & 11 12 13.4 & -61 17 58.5 &  -   &  -   &  -   &  -   &  -   &  -   & 10.0 & 0.06 & e\\
237 & 11 12 14.0 & -61 15 29.4 &  -   &  -   &  -   &  -   &  -   &  -   & 10.6 & 0.10 & e\\
238 & 11 12 16.1 & -61 16 11.4 &  -   &  -   &  -   &  -   &  -   &  -   & 09.8 & 0.06 & e\\
239 & 11 12 16.8 & -61 16 16.2 &  -   &  -   &  -   &  -   &  -   &  -   & 10.5 & 0.08 & e\\
240 & 11 12 18.0 & -61 15 42.7 &  -   &  -   &  -   &  -   &  -   &  -   & 10.7 & 0.10 & e\\
241 & 11 12 18.2 & -61 20 58.2 &  -   &  -   &  -   &  -   &  -   &  -   & 10.4 & 0.05 & e\\
242 & 11 12 22.0 & -61 20 52.5 &  -   &  -   &  -   &  -   &  -   &  -   & 11.1 & 0.07 & e\\
243 & 11 12 26.6 & -61 20 17.7 &  -   &  -   &  -   &  -   &  -   &  -   & 10.6 & 0.04 & e\\
244 & 11 12 28.5 & -61 22 21.2 &  -   &  -   &  -   &  -   &  -   &  -   & 09.6 & 0.03 & e\\
245 & 11 12 29.4 & -61 19 55.6 &  -   &  -   &  -   &  -   &  -   &  -   & 11.0 & 0.05 & e\\
246 & 11 12 29.7 & -61 20 27.9 &  -   &  -   &  -   &  -   &  -   &  -   & 10.8 & 0.06 & e\\
247 & 11 12 30.0 & -61 21 29.7 &  -   &  -   &  -   &  -   &  -   &  -   & 09.7 & 0.04 & e\\
248 & 11 12 30.8 & -61 21 19.8 &  -   &  -   &  -   &  -   &  -   &  -   & 10.3 & 0.04 & e\\
249 & 11 12 32.9 & -61 20 14.7 &  -   &  -   &  -   &  -   &  -   &  -   & 10.9 & 0.04 & e\\
250 & 11 12 41.6 & -61 16 06.0 &  -   &  -   &  -   &  -   &  -   &  -   & 11.1 & 0.05 & e\\
251 & 11 12 55.4 & -61 16 43.5 &  -   &  -   &  -   &  -   &  -   &  -   & 09.6 & 0.24 & e\\
\end{longtable}


\begin{thebibliography}{}

  \bibitem[1973]{allen} Allen, C.W. 1973,
      Astrophysical Quantities, 3rd ed.,
      (The Athlone Press, London)


  \bibitem[2003]{barbosaetal} Barbosa, C.L., Daminelli, A., Blum, R.D., Conti, P.S.
     2003, AJ, 126, 2411

   \bibitem[2001]{behrendmaeder} Behrend, R., \& Maeder, A.
     2001, A\&A, 373, 190

   \bibitem[2000]{burtonetal} Burton, M.G., Ashley, M.C.B., Marks, R.D., et al.
     2000, ApJ, 542, 359

   \bibitem[1989]{caswelletal89} Caswell, J.L., Batchelor, R.A., Foster, J.R., Wellington, K.S.
     1989, Austral. J. Phys., 42, 331

   \bibitem[1995]{caswelletal95} Caswell, J.L., Vaile, R.A., Ellingsen, S.P., Whiteoak, J.B., Norris, R.P.
     1995, MNRAS, 272, 96

   \bibitem[2003]{cutrietal} Cutri, R.M., Skrutskie, M.F., van Dyk, S., et al.
     2003, VizieR On-line Data Catalogue:II/246. Originally published in: University of Massachusetts
     and Infrared Processing and Analysis Center, (IPAC/California Institute of Technology)

   \bibitem[1999]{depreeetal} DePree, C.G., Nysewander, M.C., Goss, W.M.
     1999, AJ, 117, 2902

   \bibitem[2002]{figueredoetal} Figuer\^edo, E., Blum, R.D., Daminelli, A., Conti, P.S.
     2002, AJ, 124, 2739

   \bibitem[1998]{fowleretal} Fowler, A.M., Sharp, N., Ball, W., et al.
     1998, Proc. SPIE Vol. 3354, 1170, Infrared Astronomical Instrumentation, ed. Fowler, A.M.

   \bibitem[1974]{frogelpersson} Frogel, J.A., \& Persson, S.E.
     1974, ApJ, 192, 351

   \bibitem[1970]{gossshaver} Goss, W.M., \& Shaver, P.A.
     1970, Austral. J. Phys. Astrophys. Suppl., 14, 1

   \bibitem[2001]{haischetal} Haisch, K.E., Jr., Lada, E.A., Lada, C.J.
     2001, ApJ, 553, L153

   \bibitem[1997]{hansonetal} Hanson, M.M., Howarth, I.D., Conti, P.S.
     1997, ApJ, 489, 698

   \bibitem[1994]{hereld} Hereld, M.
     1994, ExA, 3, 87

   \bibitem[1994]{hollenbachetal} Hollenbach, D., Johnstone, D., Lizano, S., Shu, F.
     1994, ApJ, 428, 654

   \bibitem[1995]{kenyonhartmann} Kenyon, S.J., \& Hartmann, L.
     1995, ApJS, 101, 117

   \bibitem[1994]{kleinmannetal} Kleinmann, S.G., Lysaght, M.G., Pughe, W.L., et al.
     1994, Ap\&SS, 217, 11
  
   \bibitem[1983]{koorneef} Koorneef, J.
     1983, A\&A, 128, 84

   \bibitem[1992]{ladaadams} Lada, C.J., \& Adams, F.C.
     1992, ApJ, 393, 278
 
   \bibitem[2000]{ladaetal} Lada, C.J., Muench, A.A., Haisch, K.E., Jr., et al.
     2000, AJ, 120, 3162

   \bibitem[2005]{maerckerburton} Maercker, M., \& Burton, M.G.
   2005, A\&A, 438, 663

   \bibitem[1968]{mcgeegardner} McGee, R.X., \& Gardner, F.F.
     1986, Austral. J. Phys. Suppl., 21, 149

   \bibitem[1984]{mcgregoretal2} McGregor, P.J., Persson, S.E., \& Geballe, T.R.
     1984, PASP, 96, 315

   \bibitem[1994]{mcgregor} McGregor, P.J., Hart, J., Downing, M., Hoadley, D., Bloxham, G.
     1994, ExA, 3, 139

   \bibitem[1992]{moneti} Moneti, A.
     1992, A\&A, 259, 627

   \bibitem[1983]{moorwoodsalinari} Moorwood, A.F.M., \& Salinari, P.
     1983, A\&A, 125, 354

   \bibitem[1986]{olivamoorwood} Oliva, E., \& Moorwood, A.F.M.
    1986, A\&A, 164, 104

   \bibitem[1994]{persietal} Persi, P., Roth, M., Tapia, M., Ferrari-Toniolo, M., Marenzi, A.R.
     1994, A\&A, 282, 474
  
   \bibitem[2003]{rathborne} Rathborne, J. 2003,
     Young Massive Stars: Traffic Lights for Nearby Star Formation,
     PhD thesis, University of New South Wales, Sydney

   \bibitem[1980]{retallackgoss} Retallack, D.S, \& Goss, W.M.
     1980, MNRAS, 193, 261

   \bibitem[1955]{salpeter} Salpeter, E.E.
     1955, ApJ, 121, 161

   \bibitem[2001]{walshetal} Walsh, A.J., Bertoldi, F., Burton, M.G., Nikola, T.
     2001, MNRAS, 326, 36

   \bibitem[1983]{whitephillips} White, G.J., \& Phillips, J.P.
     1983, MNRAS, 202, 255

   \bibitem[1970]{wilsonetal} Wilson, T.L., Mezger, P.G., Gradner, F.F., Milne, D.K.
     1970, A\&A, 6, 364

\end{thebibliography}
\end{document}